\documentclass[aps,prd,preprintnumbers,groupedaddress,nofootinbib,amssymb,eqsecnum,notitlepage]{revtex4-2}
\usepackage{here}
\usepackage[dvipdfmx]{graphicx}
\usepackage{amsmath,amsthm,amssymb}
\usepackage{bm}
\usepackage{color}

\usepackage{amsfonts}
\usepackage{dcolumn}
\usepackage{booktabs}
\usepackage[bookmarks=true,bookmarksnumbered=true,colorlinks=false]{hyperref}
\allowdisplaybreaks[1]
\usepackage{stackengine}

\begin{document}
\newcommand{\newc}{\newcommand}

\newc{\rk}[1]{{\color{red} #1}}
\newc{\ben}{\begin{eqnarray}}
\newc{\een}{\end{eqnarray}}
\newc{\be}{\begin{equation}}
\newc{\ee}{\end{equation}}
\newc{\ba}{\begin{eqnarray}}
\newc{\ea}{\end{eqnarray}}
\newc{\D}{\partial}
\newc{\rH}{{\rm H}}
\newc{\vp}{\varphi}
\newc{\rd}{{\rm d}}
\newc{\pa}{\partial}
\newc{\Mpl}{M_{\rm Pl}}

\newcommand{\ma}[1]{\textcolor{magenta}{#1}}
\newcommand{\cy}[1]{\textcolor{cyan}{#1}}
\newcommand{\mm}[1]{\textcolor{blue}{[MM:~#1]}}
\newcommand{\re}[1]{\textcolor{red}{#1}}

\definecolor{mygreen}{RGB}{0,130,0} 
\newcommand{\red}[1]{\textcolor{red}{#1}} % 対処が必要な部分
\newcommand{\blue}[1]{\textcolor{blue}{#1}} % 対処して確認待ちの部分
\newcommand{\green}[1]{\textcolor{mygreen}{#1}} % 本文ではないコメント
\newcommand{\ora}[1]{\textcolor{orange}{#1}}

\newcommand{\an}[1]{\textcolor{mygreen}{AN: #1}} % AN のコメント

\begin{flushright}
WUCG-23-11 \\
\end{flushright}

\title{Gravitational-wave constraints on scalar-tensor gravity \\
from a neutron star and black-hole binary GW200115}

\author{Hiroki Takeda$^{1}$, Shinji Tsujikawa$^{2}$, 
and Atsushi Nishizawa$^{3}$}

\affiliation{
$^1$Department of Physics, Kyoto University, Kyoto 606-8502, Japan\\
$^2$Department of Physics, Waseda University, 3-4-1 Okubo, Shinjuku, Tokyo 169-8555, Japan\\
$^3$Physics Program, Graduate School of Advanced Science and Engineering, Hiroshima University, Higashi-Hiroshima, Hiroshima 739-8526, Japan}

\begin{abstract}

In nonminimally coupled theories where a scalar field $\phi$ 
is coupled to the Ricci scalar, neutron stars (NSs) can have 
scalar charges through an interaction with matter mediated 
by gravity. On the other hand, the same theories do not give 
rise to hairy black hole (BH) solutions. 
The observations of gravitational waves (GWs) emitted from 
an inspiralling NS-BH binary system allows a possibility of 
constraining the NS scalar charge.
Moreover, the nonminimally coupled scalar-tensor theories generate 
a breathing scalar mode besides two tensor polarizations.
Using the GW200115 data of the coalescence of 
a BH-NS binary, we place observational constraints on 
the NS scalar charge as well as the nonminimal 
coupling strength for a subclass of massless Horndeski 
theories with a luminal GW propagation. 
Unlike past related works, we exploit a waveform 
for a mixture of tensor and scalar polarizations. 
Taking the breathing mode into account, the scalar charge is more 
tightly constrained in comparison to the analysis of the tensor 
GWs alone. In nonminimally coupled theories including Brans-Dicke gravity 
and spontaneous scalarization scenarios with/without a kinetic screening, 
we put new bounds on model parameters of each theory.
\end{abstract}
\date{\today}

%\pacs{04.50.Kd, 95.36.+x, 98.80.-k}

\maketitle

%%%%%%%%%%%%%%%%%%%%%%%%
\section{Introduction}
\label{introsec}
%%%%%%%%%%%%%%%%%%%%%%%%

The direct detection of gravitational waves (GWs) emitted during 
the merger of a binary black hole (BH) opened up 
a new window for probing the physics in extreme gravity 
regimes \cite{LIGOScientific:2016aoc}. 
The first discovery of GWs has been followed by a wealthy 
of compact binary events including neutron star (NS) mergers~\cite{LIGOScientific:2021djp}.  
In particular, the NS-NS merger event GW170817 \cite{LIGOScientific:2017vwq}, along with an electromagnetic counterpart \cite{Goldstein:2017mmi}, 
showed that the speed of gravity is 
very close to that of light \cite{LIGOScientific:2017zic}. 
The same GW event offered an interesting possibility of constraining 
the matter equation of state (EOS) through the tidal deformation of NSs.
Moreover, the coalescence of a BH-NS binary was 
detected as the GW200115 event \cite{LIGOScientific:2021qlt}, 
which is also useful to test the physics in strong gravity regimes further.

General Relativity (GR) is a fundamental theory of gravity 
consistent with solar-system constraints \cite{Will:2014kxa} 
and Earth laboratory tests with high degrees of precision~\cite{Hoyle:2000cv,Adelberger:2003zx}. 
With gravitational waves from compact binary coalescences observed by LIGO-Virgo-KAGRA (LVK) collaboration, the tests of GR in the strong gravitational fields have also been actively performed~\cite{LIGOScientific:2021sio}. From the cosmological side, there are the long-standing   
problems of dark matter and dark energy in the framework 
of GR and standard model of 
particle physics \cite{Bertone:2004pz,Copeland:2006wr}. 
To resolve these problems, one typically introduces 
additional degrees of freedom (DOFs) like a scalar field 
or a vector field \cite{DeFelice:2010aj,Clifton:2011jh,Joyce:2014kja,Koyama:2015vza,Ishak:2018his,Heisenberg:2018vsk}. 
If these new DOFs work as the sources 
for dark components in the Universe, they may also play 
some roles for the physical phenomena in the vicinity of 
BHs and NSs, which can be accessed by the analysis of GWs from compact binary coalescences. 

In GR with a minimally coupled scalar field, it is known 
that static and spherically symmetric vacuum BHs  
do not acquire an additional scalar hair \cite{Hawking:1971vc,Bekenstein:1972ny}. 
This situation is unchanged even with a scalar field $\phi$ 
nonminimally coupled to a Ricci scalar $R$ of the 
form $F(\phi)R$ \cite{Hawking:1972qk,Bekenstein:1995un,Sotiriou:2011dz,Faraoni:2017ock}, 
where $F$ is a function of $\phi$. 
In the case of NSs, the presence of matter inside the star 
gives rise to a nonvanishing value of $R$ proportional to 
the matter trace $T$. Then, the scalar field and matter 
interacts with each other through the gravity-mediated 
nonminimal coupling $F(\phi)R$. 
In this case, the scalar field can have nontrivial profiles 
in the vicinity of NSs. The background geometry is also 
modified by the coupling between the scalar field and gravity. 
Thus, the nonminimal coupling leads to the existence of hairy 
NS solutions carrying a scalar charge, while this is 
not the case for BHs.

One of the representative nonminimally coupled theories is the
so-called Brans-Dicke (BD) theory \cite{Brans:1961sx} 
described by the scalar coupling $e^{-2Q \phi/\Mpl}R$ 
with the Ricci scalar, where $\Mpl$ is the reduced Planck mass. 
The coupling constant $Q$ is 
related to the BD parameter $\omega_{\rm BD}$ according to 
the relation $Q^2=[2(3+2\omega_{\rm BD})]^{-1}$ \cite{Khoury:2003rn,Tsujikawa:2008uc,DeFelice:2010aj}.
The lowest-order four-dimensional effective action in string 
theory with a dilaton field $\phi$ \cite{Fradkin:1984pq,Callan:1985ia,Gasperini:1992em} 
corresponds to a specific case of BD theories 
with $\omega_{\rm BD}=-1$, i.e., $Q^2=1/2$.
In low-energy effective string theory, there are also 
higher-order $\alpha'$ corrections of the form $\mu(\phi)X^2$ \cite{Metsaev:1987zx,Armendariz-Picon:1999hyi}, 
where $\mu$ is a function of $\phi$ and 
$X=-\partial_{\mu}\phi \partial^{\mu}\phi/2$.
This belongs to a class of nonminimally coupled k-essence 
theories described by the Lagrangian 
${\cal L}=K(\phi,X)+F(\phi)R$ \cite{Chiba:1999ka,Armendariz-Picon:2000nqq,Tsujikawa:2007gd}.
We also note that $f(R)$ gravity \cite{Starobinsky:1980te,DeFelice:2010aj} 
belongs to a subclass of extended massive BD theories with the 
BD parameter $\omega_{\rm BD}=0$, 
i.e. $Q^2=1/6$ \cite{OHanlon:1972xqa,Chiba:2003ir,Kase:2019dqc}. 
All of these generalized BD theories allow the existence of 
hairy NS solutions.

BD theories also give rise to scalar hairs for weak gravitational 
objects like the Sun or Earth. Since there is the propagation of 
fifth forces in this case, the nonminimal coupling constant is 
constrained to be $|Q| \le 2.5 \times 10^{-3}$ for massless BD 
theories \cite{Bertotti:2003rm,Khoury:2003rn,Tsujikawa:2008uc,DeFelice:2010aj}.
To evade such tight constraints on $|Q|$, we need to resort to 
some screening mechanisms like those based on a massive scalar 
field \cite{Khoury:2003aq,Khoury:2003rn} or a Galileon-type 
derivative self-interaction \cite{Nicolis:2008in,Deffayet:2009wt,Burrage:2010rs,DeFelice:2011th,Kimura:2011dc,Koyama:2013paa}. 
On the other hand, for the nonminimal coupling with even 
power-law functions of $\phi$, there are in general 
two branches of the scalar-field profile on the static and 
spherically symmetric stars with the radial coordinate $r$: 
(i) hairy solution with $\phi'(r) \neq 0$, and 
(ii) GR solution with $\phi'(r)=0$.
On weak gravitational backgrounds, the solution can stay 
near the GR branch (ii) to evade the fifth-force constraints. 
In the vicinity of strong gravitational objects like NSs, 
the GR branch (ii) can be unstable to trigger 
tachyonic instabilities 
toward the hairy branch (i) \cite{Damour:1993hw,Damour:1996ke}. 
This is a nonperturbative phenomenon known as spontaneous 
scalarization. For the nonminimal coupling 
$F(\phi)=e^{-\beta \phi^2/(2\Mpl^2)}$ advocated by 
Damour and Esposite-Farese (DEF) \cite{Damour:1993hw}, 
spontaneous scalarization of NSs occurs for a negative coupling 
constant in the range
$\beta \le -4.35$ \cite{Harada:1998ge,Novak:1998rk,Silva:2014fca,Barausse:2012da} 
(see Refs.~\cite{Shibata:2013pra,Shao:2017gwu} for the dependence 
of the upper limit of $\beta$ on the NS EOSs).
In such cases the NSs can have large scalar 
charges, while the local gravity constraints are trivially satisfied.

The binary system containing compact objects with scalar hairs 
emits scalar radiation besides tensor radiation during the merging process.
From binary pulsar measurements of the energy loss through 
the dipolar scalar radiation, the coupling $\beta$ in the 
DEF model for the scalarized NS is constrained to be 
$\beta \geq -4.5$ \cite{Freire:2012mg,Shao:2017gwu} 
(see also Ref.~\cite{Zhao:2022vig} for latest constraints).
Since the scalar radiation emitted during the inspiral phase of 
binaries also modifies the gravitational waveform, 
it is possible to derive independent constraints on the scalar 
charge and model parameters of theories.
In this vein, the gravitational waveforms in nonminimally coupled 
scalar-tensor theories have been computed in 
Refs.~\cite{Eardley1975,Will:1994fb,Alsing:2011er,Berti:2012bp,Lang:2013fna,Mirshekari:2013vb,Lang:2014osa,Sennett:2016klh,Sagunski:2017nzb,Bernard:2018hta,Liu:2020moh,Bernard:2022noq,Higashino:2022izi} to probe the deviation from 
GR through the GW observations (see also Refs.~\cite{Shibata:1994qd,Harada:1996wt,Brunetti:1998cc,Berti:2004bd,Scharre:2001hn,Chatziioannou:2012rf,Zhang:2017srh,Liu:2018sia,Niu:2019ywx}). 

If we restrict scalar-tensor theories to those with second-order field 
equations of motion and with the luminal GW propagation, 
the Lagrangian is constrained to be of the form 
${\cal L}=G_2(\phi, X)-G_3(\phi, X) \square \phi+G_4(\phi)R$, 
where $G_2, G_3$ depend on $\phi, X$ and $G_4$ is a function 
of $\phi$ alone \cite{Kobayashi:2011nu,DeFelice:2011bh,Kase:2018aps}. 
This Lagrangian, which belongs to a subclass of Horndeski 
theories \cite{Horndeski:1974wa}, accommodates all the classes 
of nonminimally coupled theories mentioned above.
On the other hand, it is known that there are no 
asymptotically-flat hairy BH solutions even with such 
general theories \cite{Hawking:1972qk,Bekenstein:1995un,Sotiriou:2011dz,Graham:2014mda,Faraoni:2017ock,Minamitsuji:2022mlv,Minamitsuji:2022vbi}.
Since the observations of gravitational waveforms from inspiralling 
compact binaries place constraints on the difference of scalar charges 
between the two objects \cite{Alsing:2011er,Yunes:2011aa,Berti:2018cxi,Tahura:2018zuq,Liu:2020moh,Higashino:2022izi}, the NS-BH binary is a most ideal system for 
extracting the information of the NS scalar charge in theories 
with the vanishing BH scalar charge.

In Ref.~\cite{Higashino:2022izi}, the inspiral gravitational waveforms 
in the above subclass of Horndeski theories were computed under 
a post-Newtonian (PN) expansion of the energy-momentum tensors of
two-point particle sources. 
For this purpose, the nonlinearity arising from the Galileon-type self interaction in $G_3(X)$ was neglected for 
the wave propagation from the source to the observer.
This amounts to imposing the condition that the Vainshtein radius $r_V$ \cite{Vainshtein:1972sx} is smaller than the size $r_s$ of compact objects. 
Besides the two tensor modes $h_+$ and $h_{\times}$, 
there are also the breathing ($h_b$) and longitudinal ($h_L$) 
polarizations arising from the scalar-field perturbation 
coupled to gravity.
The scalar radiation emitted during the inspiral phase modifies 
the phases and amplitudes of tensor GWs. 
In particular, the difference of scalar charges appears 
at the $-1$PN order in the phases of all polarizations. 
This allows us to put tight constraints on the NS scalar charge 
from observations of the NS-BH binary system. 
While the amplitudes of scalar GWs are generally 
suppressed relative to those of tensor GWs~\cite{Takeda:2023mhl},
they can provide additional observational bounds 
on the model parameters of scalar-tensor theories. 

In this paper, we will perform a test of alternative theories of 
gravity with the observational data of the NS-BH binary event 
GW200115 \cite{LIGOScientific:2021qlt} to place constraints 
on the hairy NSs realized by the subclass of Horndeski theories 
mentioned above. We focus on massless theories with the vanishing 
scalar-field mass ($m_s=0$), in which case there are three 
polarized waves ($h_+$, $h_{\times}$, $h_b$). 
In Ref.~\cite{Higashino:2022izi} the gravitational waveforms 
in the frequency domain were computed only for the tensor 
modes $h_+$ and $h_{\times}$, so we will also derive a 
frequency-domain waveform of the breathing mode $h_b$ in this paper.
We perform a statistical analysis using a complete waveform model in a subclass of Horndeski theory that includes both tensor and scalar modes, evaluate parameter correlations, and demonstrate that information independent of the phase evolution of tensor modes can be drawn from scalar 
GW amplitudes. 
A similar analysis with the GW200115 data was carried out 
in Ref.~\cite{Niu:2021nic} for generalized BD theories, but it is 
based on the waveform of tensor modes alone. 
We also note that forecast constraints on the scalar charge with Advanced LIGO and 
Einstein Telescope were studied in Ref.~\cite{Quartin:2023tpl} 
by resorting to the tensor waveform. 
Since lack of waveform elements can cause parameter bias and misinterpret constraints on the theory, one needs to be careful when interpreting test of GR results for phenomenological probes of some deviation from GR with a specific theory. The presence of scalar GWs is a key feature of 
nonminimally coupled scalar-tensor theories, so 
it is important to implement such a new polarized mode in the analysis. 
We will put bounds on the scalar charge and the nonminimal coupling strength in a more general class of theories studied in Ref.~\cite{Niu:2021nic} and then provide constraints on the allowed parameter space of each theory.

%%%%%%%%%%%%%%%%%%%%%%%%%%%%%%%%%%%%%%%%%%%%%%%%%%%%%%%%%%%%%%%%%%
\section{Gravitational waveforms in nonminimally coupled theories}
\label{sec:theory}
%%%%%%%%%%%%%%%%%%%%%%%%%%%%%%%%%%%%%%%%%%%%%%%%%%%%%%%%%%%%%%%%%%

We first revisit the tensor gravitational waveforms for a 
general class of scalar-tensor theories 
derived in Ref.~\cite{Higashino:2022izi}. 
Then, we obtain the scalar waveform in the frequency domain by 
taking into account the effect of energy loss through the 
tensor and scalar radiations. 
The most general class of scalar-tensor theories 
with second-order field equations of motion is known as 
Horndeski theories \cite{Horndeski:1974wa}. 
If we further demand that the speed of tensor GWs is exactly 
equivalent to that of light on an isotropic cosmological 
background, the Horndeski's action is restricted to be 
of the form \cite{Kobayashi:2011nu,DeFelice:2011bh,Kase:2018aps}
\be
{\cal S}=
\int {\rm d}^4 x \sqrt{-g} 
\left[ G_2(\phi,X)-G_3(\phi,X) \square \phi
+G_4(\phi)R \right]+{\cal S}_m(g_{\mu \nu}, \Psi_m)\,,
\label{action}
\ee
where $g$ is a determinant of the metric tensor $g_{\mu\nu}$, 
$G_2$ and $G_3$ are functions of $\phi$ and 
$X=-g^{\mu \nu}\partial_{\mu}\phi \partial_{\nu}\phi/2$, 
$G_4$ is a function of $\phi$ alone, 
and ${\cal S}_m$ is the action of matter fields $\Psi_m$ 
minimally coupled to gravity.

As we mentioned in Introduction, the action (\ref{action}) 
can encompass a wide class of scalar-tensor theories 
listed below.
\begin{itemize}
\item (1)~BD theories:
\be
G_2=(1-6Q^2)F(\phi)X\,,\qquad G_3=0\,,\qquad 
G_4=\frac{\Mpl^2}{2}F(\phi)\,,
\label{model1}
\ee
with the nonminimal coupling 
\be
F(\phi)=e^{-2Q\phi/\Mpl}\,,
\label{expo}
\ee
where $\Mpl$ is the reduced Plack mass. 
This is equivalent to the original BD theory  \cite{Brans:1961sx} 
with the correspondence 
$Q^2=[2(3+2\omega_{\rm BD})]^{-1}$, 
where $\omega_{\rm BD}$ is a BD parameter \cite{Khoury:2003rn,Tsujikawa:2008uc,DeFelice:2010aj}. 
GR corresponds to the limit $\omega_{\rm BD} \to \infty$, 
i.e., $Q \to 0$. 
The lowest-order effective action in string theory with a dilaton 
field $\phi$ \cite{Fradkin:1984pq,Callan:1985ia,Gasperini:1992em} 
corresponds to the specific case of BD theories 
with $\omega_{\rm BD}=0$.
In extended massive BD theories, the scalar potential 
$V(\phi)$ is present as the form $-V(\phi)$ in $G_2$.
The $f(R)$ gravity is a special case of extended massive BD theories 
with the BD parameter $\omega_{\rm BD}=0$ \cite{OHanlon:1972xqa,Chiba:2003ir}.
\item (2)~Theories of spontaneous scalarization of NSs 
with a higher-order kinetic term:
\be
G_2=\left( 1-\frac{3\Mpl^2F_{,\phi}^2}{2F^2} \right) 
F(\phi)X+\mu(\phi)X^2\,,\qquad 
G_4=\frac{\Mpl^2}{2}F(\phi)\,,
\label{model2}
\ee
where $F$ is an even power-law function of $\phi$. 
The typical example of the nonminimal coupling is 
of the form \cite{Damour:1993hw}
\be
F(\phi)=e^{-\beta \phi^2/(2\Mpl^2)}\,.
\label{spon}
\ee
The higher-order kinetic Lagrangian $\mu(\phi)X^2$, where 
$\mu$ is a function of $\phi$, belongs to the k-essence Lagrangian. 
This term allows the possibility of suppressing the NS scalar charge 
in comparison to the original spontaneous 
scalarization scenario \cite{Higashino:2022izi}.
We also note that, for the string dilaton with the nonminimal coupling 
(\ref{expo}), the similar higher-order kinetic term arises as 
an $\alpha'$ correction \cite{Metsaev:1987zx,Armendariz-Picon:1999hyi}.
In such a case, we just need to add the contribution $\mu(\phi)X^2$ 
to the Lagrangian of BD theories. 
\item (3)~Cubic Galileons with nonminial couplings:
\be
G_2=\left( 1-\frac{3\Mpl^2F_{,\phi}^2}{2F^2} \right) 
F(\phi)X\,,\qquad G_3=\alpha_3 X\,,\qquad
G_4=\frac{\Mpl^2}{2}F(\phi)\,,
\label{model3}
\ee
where $\alpha_3$ is a constant, and $F(\phi)$ 
can be chosen as Eq.~(\ref{expo}) or (\ref{spon}). 
In the vicinity of matter sources, the cubic Galileon 
Lagrangian $\alpha_3 X \square \phi$ 
can screen fifth forces mediated by the nonminimal coupling.
This is due to the dominance of scalar-field non-linearities
within a Vainshtein radius $r_V$. 
If $r_V$ is much larger than the size of local objects $r_s$, then 
the linear expansion of scalar GWs propagating on the Minkowski 
background loses its validity for the distance $r_s<r<r_V$. 
To avoid the dominance of 
non-linearities outside the matter source, we require that 
$r_V \lesssim r_s$. Then the screening of fifth forces 
occurs inside the object, which suppresses the scalar charge. 
Since this situation is analogous to the kinetic screening induced by 
the term $\mu(\phi)X^2$ in Eq.~(\ref{model2}), we will not 
place observational constraints on theories given 
by the functions (\ref{model3}). 
We note that scalar gravitational radiation from binary 
pulsars in cubic Galileon theories was studied in 
Ref.~\cite{deRham:2012fw} and was constrained in 
Ref.~\cite{Shao:2020fka}.
\end{itemize}
In the above theories there are hairy NS solutions with 
scalar hairs, while the BHs do not have hairy 
solutions \cite{Hawking:1972qk,Bekenstein:1995un,Sotiriou:2011dz,Graham:2014mda,Faraoni:2017ock,Minamitsuji:2022mlv,Minamitsuji:2022vbi}\footnote{If we consider 
full Horndeski theories, the scalar field coupled to a Gauss-Bonnet term 
gives rise to asymptotically-flat hairy BH 
solutions \cite{Kanti:1995vq,Doneva:2017bvd,Silva:2017uqg,Antoniou:2017acq}.}.
Thus the GWs emitted from the NS-BH binary system 
can provide 
constraints on the NS scalar charge and the nonminimal coupling strength. 
We would like to translate them to the bounds on 
model parameters in each theory.

By the end of this section, we will revisit the gravitational waveforms from inspiralling compact binaries already discussed elsewhere
in scalar-tensor theories \cite{Alsing:2011er,Yunes:2011aa,Berti:2018cxi,Tahura:2018zuq,Liu:2020moh,Higashino:2022izi}. The derivation of them is mostly based on 
the reference \cite{Higashino:2022izi}, but, in this paper, we will 
newly obtain the frequency-domain gravitational waveform of 
a breathing scalar mode in Sec.~\ref{fdomainsec}. 
In Sec.~\ref{prosec}, we also take into account the cosmological 
propagation of GWs from the source to the observer.
While a similar study was performed in Ref.~\cite{Quartin:2023tpl}, 
we will provide the complete frequency-domain gravitational waveforms 
emitted from the NS-BH binary including the breathing polarization.

We deal with the NS-BH binary system as a collection 
of two point-like particles (with the label $I=A$
for NS and $I=B$ for BH). 
The matter action for such a system is given by \cite{Eardley1975}
\be
{\cal S}_m=-\sum_{I=A,B} \int m_I(\phi) {\rm d} \tau_I\,,
\label{matter}
\ee
where $m_I(\phi)$'s are the $\phi$-dependent ADM masses of 
compact objects, and $\tau_I$ is the proper time along a world 
line $x_{I}^{\mu}$ of the particle $I$. 
The matter energy-momentum tensor $T^{\mu \nu}$ follows from 
the variation of ${\cal S}_m$ with respect to $g_{\mu \nu}$, 
as $\delta {\cal S}_m=(1/2)\int \rd^4 x \sqrt{-g}\,T^{\mu \nu} 
\delta g_{\mu \nu}$. In terms of the matter trace 
$T=g_{\mu \nu}T^{\mu \nu}$, the action (\ref{matter}) 
can be expressed as ${\cal S}_m=\int \rd^4 x \sqrt{-g}\,T(\phi)$. 
More explicitly, the trace $T$ is related to the $\phi$-dependent 
masses of sources, as \cite{Eardley1975,Higashino:2022izi}
\be
T(\phi)=-\frac{1}{\sqrt{-g}} 
\sum_{I=A,B} m_I(\phi) 
\frac{1}{u_I^0}\delta^{(3)} ({\bm x}-{\bm x}_I(t))\,,
\label{trace}
\ee
where $u_I^0$ is the time component of four velocity of the particle $I$, 
and $\delta^{(3)} ({\bm x}-{\bm x}_I(t))$ is the three dimensional 
delta function with the spatial particle position ${\bm x}_I(t)$ 
at time $t$.

\subsection{Solutions to tensor and scalar waves}

In the following, we study the propagation of GWs from the binary 
to an observer in the subclass of Horndeski theories 
given by the action (\ref{action}). 
We consider metric perturbations $h_{\mu \nu}$ on a 
Minkowski background with the metric tensor 
$\eta_{\mu \nu}={\rm diag} (-1,1,1,1)$, such that 
\be
g_{\mu \nu}=\eta_{\mu \nu}+h_{\mu \nu}\,.
\ee
The scalar field $\phi$ is expanded around today's 
constant asymptotic value $\phi_0$, as
\be
\phi=\phi_0+\varphi\,,
\ee
where $\varphi$ corresponds to a perturbed quantity.
The background scalar $\phi_0$ is determined by 
the cosmological evolution from the past to today, 
which will be discussed in Sec.~\ref{prosec}.

For the later convenience, we introduce 
the following combination 
\be
\theta_{\mu \nu} \equiv h_{\mu \nu}-\frac{1}{2} h \eta_{\mu \nu}
-\xi_0 \frac{\varphi}{\Mpl} \eta_{\mu \nu}\,,
\label{thetamn}
\ee
where $h$ is the trace of $h_{\mu \nu}$, and $\xi_0$ is 
defined by 
\be
\xi_0 \equiv \frac{\Mpl G_{4,\phi}}{G_4} \biggl|_{\phi=\phi_0}\,,
\label{xi_0}
\ee
with the notation $G_{4,\phi}=\rd G_4/\rd \phi$. 
Choosing the Lorentz-gauge condition $\partial^{\nu} \theta_{\mu \nu}=0$, 
the perturbation $\theta_{\mu \nu}$ obeys
\be
\square_{\rm M} \theta_{\mu \nu}=-\frac{T_{\mu \nu}^{(1)}}{G_4(\phi)}
+{\cal O} \left(\theta^2, \vp^2, \theta \vp, T_{\mu \nu}^{(2)},\cdots 
\right)\,,
\label{thetamu}
\ee
where $\square_{\rm M} \equiv \eta^{\mu \nu} \partial_{\mu} \partial_{\nu}$, 
$T_{\mu \nu}^{(1)}$ and $T_{\mu \nu}^{(2)}$ are the first- and 
second-order perturbations of $T_{\mu \nu}$ respectively, 
and $\theta$ is the trace of $\theta_{\mu \nu}$. 
The equation of motion for the scalar-field 
perturbation is given by 
\ba
\left( \square_{\rm M}-m_s^2 \right) \vp
&=& 
-\frac{1}{\zeta_0} \left( 1-\frac{1}{2}\theta
-\xi_0 \frac{\varphi}{\Mpl}-\frac{\zeta_1}{\zeta_0} 
\varphi \right) \left( T_{,\phi}-\frac{G_{4,\phi}}{2G_4} 
T \right) \nonumber\\
& & +{\cal O} \left( \varphi^2, \partial_{\mu} \varphi 
\partial^{\mu} \varphi, (\square_{\rm M} \varphi)^2, 
\partial^{\mu}\partial^{\nu} \varphi
\partial_{\mu}\partial_{\nu} \varphi, 
\theta \varphi, 
\theta^{\mu \nu} \partial_{\mu} \partial_{\nu} \varphi \right)\,,
\label{vpeqf}
\ea
where 
\be
\zeta_0 \equiv G_{2,X}
+\frac{3G_{4,\phi}^2}{G_4} \biggr|_{\phi=\phi_0}\,,\qquad 
\zeta_1 \equiv
G_{2,\phi X}
+\frac{6 G_{4,\phi} G_{4,\phi \phi}}{G_4}
-\frac{3G_{4,\phi}^3}{(G_4)^2} \biggr|_{\phi=\phi_0}\,,\qquad
m_s^2 \equiv -\frac{G_{2,\phi \phi}(\phi_0)}{\zeta_0}\,.
\label{ze0def}
\ee
The quantity $m_s$ corresponds to the scalar-field mass.  
In theories given by the Horndeski functions (\ref{model1}), 
(\ref{model2}), and (\ref{model3}), we have $G_{2,\phi \phi}=0$ 
by using the property $X|_{\phi=\phi_0}=0$.
Then, we have 
\be
m_s=0\,.
\label{massless}
\ee
In the following, we will focus on massless theories satisfying 
the condition (\ref{massless}).
In this case, the scalar GWs arising from the field perturbation 
$\varphi$ have only the breathing polarization, 
without the longitudinal propagation.

Since the right-hand sides of Eqs.~(\ref{thetamu}) and (\ref{vpeqf}) 
contain the $\phi$-dependent quantities $G_4(\phi)$ and 
$m_{I}(\phi)$, we expand them around $\phi=\phi_0$ as 
\be
G_4(\phi)=G_4(\phi_0) \left[ 1+\xi_0 \frac{\varphi}{\Mpl}
+{\cal O}(\varphi^2) \right]\,,\qquad 
m_I(\phi)=m_I(\phi_0) \left[ 1+\alpha_I \frac{\varphi}{\Mpl}
+{\cal O}(\varphi^2) \right]\,,
\label{G4expan}
\ee
where 
\be
\alpha_I \equiv \frac{\Mpl m_{I,\phi}}{m_I} 
\biggl|_{\phi=\phi_0}\,.
\label{alphaI}
\ee

In the Jordan-frame action (\ref{action}), the scalar field $\phi$ 
is directly coupled to the Ricci scalar $R$. 
Performing a conformal transformation 
$\hat{g}_{\mu \nu}=[2G_4(\phi)/\Mpl^2] g_{\mu \nu}$ 
of the metric, we obtain the action where the gravitational 
sector is described by the Einstein-Hilbert term 
$\Mpl^2 \hat{R}/2$, where a hat represents 
quantities in the transformed Einstein frame.
In the Einstein frame, the 
scalar field is coupled to matter fields through the 
metric tensor $g_{\mu \nu}=[2G_4(\phi)/\Mpl^2]^{-1}\hat{g}_{\mu \nu}$.
If we consider a star on a spherically symmetric background, 
the scalar field can acquire a charge $q_s$ through 
the interaction with matter mediated by the nonminimal coupling. 
Provided that the kinetic term $\hat{X}$ is the dominant 
contribution to the scalar-field Lagrangian in the Einstein 
frame at large radial distance $\hat{r}$, the field has 
the following asymptotic behavior 
\be
\phi(\hat{r})=\phi_0-\frac{q_s}{\hat{r}}\,,
\label{phiL}
\ee
whose radial derivative $\phi'(\hat{r})=q_s/\hat{r}^2$ 
is analogous to the electric field in electrodynamics.
For the star ADM mass $\hat{m}_I (\phi)$ in the Einstein frame, 
the scalar charge $q_s$ has a relation with the $\phi$ derivative of 
$\hat{m}_I (\phi)$ as $4 \pi q_s=\hat{m}_{I,\phi}$ \cite{Damour:1992we}.
In the Einstein frame, we introduce a dimensional quantity 
analogous to Eq.~(\ref{alphaI}) as 
\be
\hat{\alpha}_I \equiv \frac{\Mpl \hat{m}_{I,\phi}}{\hat{m}_I} 
\biggl|_{\phi=\phi_0}=\frac{4\pi \Mpl}{\hat{m}_I}q_s\,.
\label{alphaIhat}
\ee
This means that $\hat{\alpha}_I$ characterises the strength of 
the scalar charge $q_s$. 
Since the star ADM mass $m_I$ in the Jordan frame is 
related to $\hat{m}_I$ as $m_I=\sqrt{2G_4(\phi)/\Mpl^2}\,\hat{m}_I$, 
we have the following relation 
\be
\hat{\alpha}_I=\alpha_I-\frac{1}{2}\xi_0\,.
\ee
The dimensionless quantity $\hat{\alpha}_I$ is more fundamental 
than $\alpha_I$ due to the direct relation with the scalar charge,  
so we will express the gravitational waveforms 
by using $\hat{\alpha}_I$.
We note that all the calculations given below will be performed 
in the Jordan frame, except for replacing $\alpha_I$ 
with $\hat{\alpha}_I+\xi_0/2$.

For the binary system, we consider a relative circular orbit   
rotating around a fixed center of mass. 
Then, the Newtonian equation along the radial direction 
is expressed as 
\be
\mu \frac{v^2}{r}=
\frac{\tilde{G} m_A m_B}{r^2}\,,
\label{gere2}
\ee
where $v$ and $r$ are the relative speed and displacement 
of two sources, respectively, and \cite{Liu:2020moh,Higashino:2022izi}
\ba
\mu  &\equiv& \frac{m_A m_B}{m}\,,\qquad 
m \equiv m_A+m_B\,,\\
\tilde{G} &\equiv& 
\frac{1}{16 \pi G_4(\phi_0)} 
\left[
1+\frac{4G_4(\phi_0)}{\zeta_0 \Mpl^2} 
\hat{\alpha}_A \hat{\alpha}_B
\right]\,.\qquad 
\label{tG}
\ea
Note that $\mu$ is the reduced mass. 
Provided that the two compact objects $A$ and $B$ have the nonvanishing 
scalar charges $\hat{\alpha}_A$ and $\hat{\alpha}_B$, 
respectively, the effective gravitational coupling 
$\tilde{G}$ is modified by the product 
$\hat{\alpha}_A \hat{\alpha}_B$.
For the NS-BH system in which the BH does not have a scalar hair, 
we have $\hat{\alpha}_B=0$ and hence 
$\tilde{G}=1/[16 \pi G_4(\phi_0)]$. 
{}From Eq.~(\ref{gere2}), we obtain the following relation 
\be
v=( \tilde{G}m \omega )^{1/3}\,,
\ee
where $\omega=v/r$ is the angular frequency.

At distance $D$ from the binary source, the leading-order 
solution to the tensor wave Eq.~(\ref{thetamu}) is given by 
\be
\theta^{ij}=\frac{\tilde{G}\mu m}{4\pi G_4(\phi_0) r D}
\left( \hat{v}^i \hat{v}^j-\hat{r}^i \hat{r}^j \right)\,,
\label{thetaijdef}
\ee
where $\hat{v}^i$ and $\hat{r}^i$ are the unit vectors 
along the relative velocity ${\bm v}$ and displacement 
${\bm r}$ of the circular orbit (with $r=|{\bm r}|$).

For the scalar-field perturbation $\varphi$, we derive the solution to 
Eq.~(\ref{vpeqf}) up to the quadrupole order in the PN  
expansion. We take a vector field from the source to the 
observer as ${\bm D}=D{\bm n}$, where ${\bm n}$ is a unit vector. 
Dropping the time-independent monopole contributions 
to $\varphi$, the scalar-field perturbation 
measured by the observer is given by \cite{Liu:2020moh,Higashino:2022izi}
\be
\varphi 
=-\frac{\mu}{4\pi \zeta_0 \Mpl D} 
\biggl[ \left( \hat{\alpha}_A-\hat{\alpha}_B \right){\bm v} \cdot {\bm n}
-\frac12 \Gamma ({\bm v}\cdot {\bm n})^2
+\frac{\Gamma}{2}\frac{\tilde{G}m}{r^3}
 ({\bm r}\cdot {\bm n})^2
\biggr] \bigg|_{t-D}\,,
\label{varphiso}
\ee
where 
\be
\Gamma \equiv -2\frac{m_B \hat{\alpha}_A
+m_A \hat{\alpha}_B}{m}\,.
\label{Gamma}
\ee
Since we are now considering the massless theories with $m_s=0$, 
there is no contribution to $\varphi$ arising from the 
longitudinal polarization. 
If both of the scalar charges $\hat{\alpha}_A$ and $\hat{\alpha}_B$ 
are zero, then we have $\varphi=0$ in Eq.~(\ref{varphiso}).
As long as either $\hat{\alpha}_A$ or $\hat{\alpha}_B$ is nonvanishing, 
the scalar-field perturbation does not vanish together with 
the tensor modes (\ref{thetaijdef}).

\subsection{Time-domain solutions}
\label{timesec}

We are now interested in the time-domain solutions to GWs 
emitted from the binary system with a relative circular motion.
In the Cartesian coordinate system $(x_1,x_2,x_3)$ 
whose origin O is fixed at the center of mass, 
we consider an observer present in the $(x_2,x_3)$ plane. 
The unit vector ${\bm n}$ from O to the observer is 
inclined from the $x_3$ axis with an angle $\iota$. 
The binary circular motion is confined on the $(x_1,x_2)$ plane, 
with a relative vector ${\bm r}$ from O.
The angle between ${\bm r}$ and the $x_1$ axis is 
given by $\Phi$, with the velocity ${\bm v}=\dot{\bm r}$ 
orthogonal to ${\bm r}$.
In this configuration, one can express 
${\bm n}$, ${\bm r}$, and ${\bm v}$ as
\be
{\bm n}=(0, \sin \iota, \cos \iota)\,,\qquad 
{\bm r}=(r\cos \Phi, r \sin \Phi, 0)\,,\qquad 
{\bm v}=(-v\sin \Phi, v\cos \Phi, 0)\,,
\ee

For the tensor wave defined by Eq.~(\ref{thetamn}), 
we choose the traceless-transverse (TT) gauge conditions 
$\theta=0$ and $\partial^j \theta_{ij}=0$. 
We consider the GWs propagating along the $x_3$ direction, 
in which case $n_{x_1}=n_{x_2}=0$ and $n_{x_3}=1$. 
For a massless scalar field, the GW field 
can be expressed as the $3 \times 3$ matrix 
components as \cite{Eardley:1973zuo,Maggiore:1999wm,Maggiore:2007ulw}
\ba
{\bm h}_{ij}=
\left(
\begin{array}{ccc}
h_{+}+h_{b} & h_{\times} & 0 \\
h_{\times}  & -h_{+}+h_{b} & 0 \\
0 & 0 & 0 \\
\end{array}
\right)\,,
\ea
where
\be
h_{+}=\theta_{11}^{\rm TT}=-\theta_{22}^{\rm TT}\,,\qquad
h_{\times}=\theta_{12}^{\rm TT}=\theta_{21}^{\rm TT}\,,\qquad
h_b=-\xi_0 \frac{\varphi}{\Mpl}\,,
\label{perturbation_to_polarization}
\ee
with $\theta_{11}^{\rm TT}$, $\theta_{22}^{\rm TT}$, 
$\theta_{12}^{\rm TT}$ being the TT components 
of $\theta_{ij}$. We have two tensor polarizations 
$h_{+}$ and $h_{\times}$ besides the breathing mode $h_b$.
Since we are now considering the 
massless theories with $m_s=0$, the longitudinal 
mode $h_{L}$ does not appear as the (33) component 
in ${\bm h}_{ij}$. 

In Ref.~\cite{Higashino:2022izi}, the authors derived 
the three components $h_{+}$, $h_{\times}$, and $h_{b}$ 
in the time domain with a constant angular frequency $\omega$.
At the observer position ${\bm x}=D{\bm n}$, 
they are given, respectively, by 
\ba
h_{+} &=&
-(1+\delta_0 )^{2/3} \frac{4(G_{*} {\cal M})^{5/3} \omega^{2/3}}{D}
\frac{1+\cos^2 \iota}{2} \cos(2\Phi)\,,\label{hp}\\
h_{\times} 
&=&-(1+\delta_0 )^{2/3} \frac{4(G_{*} {\cal M})^{5/3} \omega^{2/3}}{D}
\cos \iota \sin(2\Phi)\,,\label{ht}\\
h_{b}
&=& \frac{\mu \xi_0}{4\pi \zeta_0 \Mpl^2 D} 
\left[ (\hat{\alpha}_A-\hat{\alpha}_B)
(\tilde{G}m \omega)^{1/3} \sin \iota \cos \Phi
-\frac{1}{2}\Gamma (\tilde{G} m \omega)^{2/3} 
\sin^2 \iota \cos(2\Phi) \right]\,,
\label{hb}
\ea
where $\Phi=\omega (t-D)$, and 
\be
\delta_0 \equiv 4\kappa_0 \hat{\alpha}_A \hat{\alpha}_B\,,
\qquad
\kappa_0 \equiv
\frac{G_4(\phi_0)}{\zeta_0 \Mpl^2}\,,\qquad
G_{*} \equiv \frac{1}{16 \pi G_4(\phi_0)}\,,\qquad
{\cal M} \equiv \mu^{3/5}m^{2/5}\,.
\label{Gstar}
\ee
The breathing polarized mode (\ref{hb}) depends on 
the quantities $\hat{\alpha}_A-\hat{\alpha}_B$ and 
$\Gamma$ defined by Eq.~(\ref{Gamma}).
If the two compact objects do not have any scalar charges, then 
$h_b$ vanishes. In other words, the detection of the 
breathing mode is a smoking gun for the presence of 
a scalar field nonminimally coupled to gravity.

In theories given by the action (\ref{action}), it is known 
that static and spherically symmetric BHs do not have 
scalar hairs. We will focus on the NS-BH binary system 
where the BH has a vanishing scalar charge. 
In this case, we have 
\be
\hat{\alpha}_B=0\,,\qquad 
\delta_0=0\,,\qquad 
\tilde{G}=G_*\,,\qquad
\Gamma=-\frac{4\eta}{1\pm \sqrt{1-4\eta}}
\hat{\alpha}_A\,,
\label{NSBHre}
\ee
where the plus and minus signs in $\Gamma$ correspond to 
the cases $m_A>m_B$ and $m_A<m_B$, respectively. 
In the following, we will exploit the relations in 
Eq.~(\ref{NSBHre}).

\subsection{Frequency-domain solutions with 
gravitational radiation}
\label{fdomainsec}

In Sec.~\ref{timesec} we assumed that $\omega$ is constant, 
but, in reality, the orbital frequency increases 
through gravitational radiation. 
The stress-energy tensor associated with gravitational 
radiation is given by \cite{Hou:2017cjy,AbhishekChowdhuri:2022ora}
\be
t_{\mu \nu}=\bigg \langle \frac{1}{2} G_4(\phi_0) 
\partial_{\mu} \theta_{\alpha \beta}^{\rm TT} 
\partial_{\nu} \theta^{\alpha \beta}_{\rm TT} 
+\zeta_0 \partial_{\mu} \varphi \partial_{\nu} \varphi
\bigg\rangle\,.
\label{taumn}
\ee
In scalar-tensor theories, the scalar radiation arising from 
the perturbation $\varphi$ contributes to $t_{\mu \nu}$ 
besides the tensor radiation associated 
with $\theta_{\mu \nu}$.
Due to the conservation of $t^{\mu \nu}$ inside a volume 
$V$, the derivative of gravitational energy 
$E_{\rm GW}=\int_V \rd^3 x\,t^{00}$ 
with respect to time $t$ yields
\be
\dot{E}_{\rm GW}=-\int_V \rd^3 x\,\partial_i t^{0i}
=-\int {\rm d}\Omega\,D^2 \left[ 
G_4(\phi_0) \langle \dot{h}_{+}^2+\dot{h}_{\times}^2 
\rangle-\zeta_0 \langle\partial_0 \varphi \partial_{D}\varphi 
\rangle\right]\,,
\label{eq:energy-loss}
\ee
where $\Omega$ is the solid angle element. 
The binary system has a mechanical energy 
\be
E=\frac{1}{2} \mu v^2-G_*\frac{\mu m}{r}
=-\frac{1}{2}\mu (G_*
m \omega)^{2/3}\,.
\ee
Since $\dot{E}=\dot{E}_{\rm GW}$, the orbital 
frequency $\omega$ increases in time. 
We substitute Eqs.~Eqs.~(\ref{varphiso}), (\ref{hp}), and 
(\ref{ht}) into 
Eq.~(\ref{eq:energy-loss}) and use 
the relation $\dot{E}=\dot{E}_{\rm GW}$ to find $\dot{\omega}$. 
This calulation was already performed in Ref.~\cite{Higashino:2022izi}.
At leading order in the PN approximation, we have
\be
\dot{\omega} \simeq 
\frac{96}{5}(G_{*} {\cal M})^{5/3} \omega^{11/3} 
\left[ 1+\frac{5 \kappa_0 \hat{\alpha}_A^2}
{24 (G_* m \omega)^{2/3}}+\frac{\kappa_0}{6}\Gamma^2 \right]\,. 
\label{dome2}
\ee

To confront the gravitational waveforms with observations, 
we perform Fourier transformations of $h_{+}$, $h_{\times}$, and $h_b$
with a frequency $f$, such 
that\footnote{Unlike Ref.~\cite{Higashino:2022izi}, we choose 
the minus sign for the phase to match it with the notation 
used later in Sec.~\ref{sec:model}.} 
\be
\tilde{h}_{\lambda} (f)=\int {\rm d}t\,h_{\lambda}(t) 
e^{-i \cdot 2 \pi f t}\,,
\ee
where $\lambda=+, \times, b$. 
Under a stationary phase approximation, 
the frequency-domain solutions to $\tilde{h}_{+}$ 
and $\tilde{h}_{\times}$ were already derived 
in Ref.~\cite{Higashino:2022izi}. 
Taking into account the quadrupole terms besides the 
dipole terms, the tensor gravitational waveformes 
are given, respectively, by 
\ba
\tilde{h}_{+} (f)
&=& -\sqrt{\frac{5\pi}{24}}\,\frac{(G_* {\cal M})^{5/6}}{D}
(\pi f)^{-7/6} 
\left[ 1-\frac{5 \eta^{2/5} \kappa_0 \hat{\alpha}_A^2}{
48(G_* {\cal M} \pi f)^{2/3}}-\frac{\kappa_0 \Gamma^2}{12}
\right] \frac{1+\cos^2 \iota}{2} e^{-i \Psi_+} \,,\label{hpf}\\
\tilde{h}_{\times}(f)
&=& -\sqrt{\frac{5\pi}{24}}\,\frac{(G_* {\cal M})^{5/6}}{D} 
(\pi f)^{-7/6} 
 \left[ 1-\frac{5 \eta^{2/5} \kappa_0 \hat{\alpha}_A^2}
 {48(G_* {\cal M} \pi f)^{2/3}}-\frac{\kappa_0 \Gamma^2}{12}
\right] (\cos \iota)\,e^{-i \Psi_\times}\,,
\label{htf}
\ea
where
\ba
& &
\Psi_{+}= 
\Psi_{\times}+\frac{\pi}{2}
= 2\pi f t_\infty -2\Phi_c-\frac{\pi}{4}
+\frac{3}{128} (G_* {\cal M} \pi f)^{-5/3}
\left[ 1-\frac{5 \eta^{2/5}\kappa_0 \hat{\alpha}_A^2}
{42 (G_* {\cal M} \pi f)^{2/3}}-\frac{\kappa_0 \Gamma^2}{6} \right]\,,
\label{Psip}\\
& &
\eta = \frac{\mu}{m}
=\left( \frac{{\cal M}}{m} \right)^{5/3}\,.
\ea
Here, $\Phi_c$ is the value of $\Phi$ at which $\omega$ 
increases to a sufficiently large value (at $t=t_\infty$). 
In the phase (\ref{Psip}), we shifted the origin of time to absorb
the distance $D$, such that $t_\infty+D \to t_\infty$. 
We also note that terms higher than the order 
$\hat{\alpha}_A^2$ are neglected for obtaining 
the results (\ref{hpf})-(\ref{Psip}).

The breathing scalar mode (\ref{hb}), which was derived for 
constant $\omega$, consists of two parts:
\ba
h_{b1}
&=& 
\frac{\mu \xi_0 \hat{\alpha}_A}
{4\pi \zeta_0 \Mpl^2 D} (G_*m \omega)^{1/3} 
\sin \iota \cos \Phi\,,\\
h_{b2}
&=&
-\frac{\mu \xi_0 \Gamma }{8\pi \zeta_0 \Mpl^2 D} 
(G_*m \omega)^{2/3} \sin^2 \iota 
\cos(2\Phi)\,.
\ea
Performing the Fourier transformation for $h_{b1}$, it follows that 
\be
\tilde{h}_{b1}(f)=\frac{\mu \xi_0 \hat{\alpha}_A}
{8\pi \zeta_0 \Mpl^2 D}  (G_*m)^{1/3} 
e^{-i \cdot 2\pi fD} \sin \iota
\int \rd t\,\omega(t)^{1/3} \left[ e^{i (\Phi(t)-2\pi ft)}
+e^{-i (\Phi(t)+2\pi ft)} \right]\,.
\label{hb1a}
\ee
The first term in the square bracket of Eq.~(\ref{hb1a}) 
has a stationary phase point characterized by 
\be
\omega(t_*)=\dot{\Phi}(t_*)=2\pi f\,.
\ee
We expand $\Phi(t)$ around $t=t_*$, as 
$\Phi(t)=\Phi(t_*)+2\pi f (t-t_*)+\dot{\omega}(t_*) (t-t_*)^2/2
+{\cal O}(t-t_*)^3$. 
Since the second term in the square bracket of Eq.~(\ref{hb1a}) 
is fast oscillating, we drop its contribution to $\tilde{h}_{b1}(f)$. 
On using the property $\int {\rm d}t\, \omega(t)^{1/3} 
e^{i \dot{\omega}(t_*) (t-t_*)^2/2} \simeq \omega(t_*)^{1/3} 
\sqrt{2\pi/\dot{\omega}(t_*)}\,e^{i \pi/4}$, we obtain 
\be
\tilde{h}_{b1}(f)=\frac{\mu \xi_0 \hat{\alpha}_A}
{8\pi \zeta_0 \Mpl^2 D} 
(G_*m)^{1/3} 
(\sin \iota)\,\omega(t_*)^{1/3} \sqrt{\frac{2\pi}
{\dot{\omega}(t_*)}}\,e^{-i \Psi_{b}}\,,
\label{hb1b}
\ee
where 
\be
\Psi_{b}=2\pi f t_{\infty}-\Phi_c-\frac{\pi}{4}
+\int_{\infty}^{2\pi f} {\rm d} \omega 
\frac{2\pi f-\omega}{\dot{\omega}}\,.
\label{hPsi}
\ee
We substitute Eq.~(\ref{dome2}) into Eqs.~(\ref{hb1b})-(\ref{hPsi}) 
and perform the integration with respect to $\omega$. 
Neglecting the terms of order $\hat{\alpha}_A^3$ in 
the amplitude of $\tilde{h}_{b1}(f)$, it follows that 
\be
\tilde{h}_{b1}(f)=\sqrt{\frac{5}{96}}
\frac{\mu \xi_0 \hat{\alpha}_A}
{16\pi \zeta_0 \Mpl^2 D} 
\frac{(G_*m)^{1/3}}{(G_* {\cal M})^{5/6}}
\frac{\sin \iota}{\pi f^{3/2}}
e^{-i \Psi_{b}}\,,
\label{hb1F}
\ee
where 
\be
\Psi_{b}=2\pi f t_{\infty}-\Phi_c-\frac{\pi}{4}
+\frac{3}{256 (2 G_* {\cal M} \pi f)^{5/3}}
\left[ 1-\frac{5 \eta^{2/5} \kappa_0 \hat{\alpha}_A^2}
{42 (2G_* {\cal M} \pi f)^{2/3}}
-\frac{\kappa_0 \Gamma^2}{6} \right]\,.
\label{hPsif}
\ee
Similarly, the Fourier-transformed mode of $h_{b2}$ 
can be derived as 
\be
\tilde{h}_{b2}(f)=-\sqrt{\frac{5}{96}}\frac{\mu \xi_0\Gamma}
{16\pi \zeta_0 \Mpl^2 D} 
 \frac{(G_*m)^{2/3}}{(G_* {\cal M})^{5/6}} 
 \frac{\sin^2 \iota}{\pi^{2/3} f^{7/6}} e^{-i \Psi_{+}}\,,
\label{hb2F}
\ee
where $\Psi_{+}$ is given by Eq.~(\ref{Psip}). 
The breathing mode $\tilde{h}_b(f)$ in the frequency domain is
the sum of Eqs.~(\ref{hb1F}) and (\ref{hb2F}). 
We will consider the asymptotic field value satisfying 
$G_4(\phi_0) \simeq \Mpl^2/2$, 
in which case $\Mpl^2 \simeq 1/(8\pi G_*)$. Then, we obtain
\be
\tilde{h}_b (f)=\tilde{h}_{b1} (f)+\tilde{h}_{b2} (f) 
\simeq 
\sqrt{\frac{5\pi}{24}}\,\frac{(G_* {\cal M})^{5/6}}{D}
(\pi f)^{-7/6}\frac{\xi_0}{4 \zeta_0} 
\biggl[ \frac{\eta^{1/5} \hat{\alpha}_A}
{(G_* {\cal M} \pi f)^{1/3}} (\sin \iota) 
e^{-i \Psi_{b}} -\Gamma\,(\sin^2 \iota) e^{-i \Psi_{+}}
\biggr]\,.
\label{hbfa}
\ee
This is a new result of the frequency-domain breathing mode, 
which was not derived in Ref.~\cite{Higashino:2022izi}.

\subsection{Inspiral GWs from NS-BH binaries 
and cosmological propagation}
\label{prosec}

So far, we have assumed that the GWs propagate on the Minkowski background. 
If the GW source is far away from the observer, the effect of 
cosmic expansion on the gravitational waveform should be 
taken into account. Let us then consider the spatially-flat 
cosmological background given by the line element
\be
{\rm d}s^2=-{\rm d}t^2+a^2(t) \delta_{ij} \rd x^i \rd x^j\,,
\ee
where $a(t)$ is a time-dependent scale factor.
The redshift of the binary source is defined by $z=a(t_0)/a(t_s)-1$, 
where $t_0$ and $t_s$ are the moments measured by the 
clocks at observer and source positions respectively. 
The GW frequency measured by 
the observer, $\tilde{f}$, is different from the one measured
in the source frame, $f$, as 
\be
\tilde{f}=(1+z)^{-1}f\,.
\ee
On the cosmological background, the time variation of $\phi$ in 
the nonminimal coupling $F(\phi)$ gives rise to a modified 
propagation of GWs. We define the luminosity distance
$d_L(z)=(1+z)\int_0^z H^{-1}(\tilde{z}){\rm d} \tilde{z}$, 
where $H=\dot{a}/a$ is the Hubble 
expansion rate. 
The effective distance $d_{\rm GW}(z)$ travelled by GWs is 
related to $d_L(z)$, as \cite{Saltas:2014dha,Nishizawa:2017nef,Belgacem:2017ihm,
Kase:2018aps,Tsujikawa:2019pih} 
\be
d_{\rm GW}(z)=d_L(z) \sqrt{\frac{G_4(\phi_0)}{G_4(\phi_s)}}\,,
\label{dGW}
\ee
where $\phi_s$ is the background scalar field when GWs
are emitted from the source.

The Lunar Laser Ranging experiment put constraints on 
the time variation of today's gravitational 
coupling $G_*=1/(16 \pi G_4)$ 
as $\dot{G}_*/G_*=(7.1\pm 7.6) \times 10^{-14}$~yr$^{-1}$ \cite{Hofmann:2018myc}, which was 
derived by assuming the evolution 
of $G_*$ linear in time.
This gives a tight bound $|\dot{G}_4/(H_0 G_4)|_{\phi=\phi_0} \lesssim 10^{-3}$ 
for general nonminimal couplings $G_4(\phi)$, 
where $H_0$ is today's Hubble expansion rate \cite{Tsujikawa:2019pih}.
Hence the time variation of $\phi$ over the cosmological 
time scale $H_0^{-1}$ is suppressed at low redshifts ($z \lesssim 1$).
Then the ratio $G_4(\phi_0)/G_4(\phi_s)$ in Eq.~(\ref{dGW}) 
can be approximated as 1, so that $d_{\rm GW}(z)$ is close to 
$d_L(z)$ for nonminimally coupled theories.

For the tensor waveforms, the analysis of Ref.~\cite{Quartin:2023tpl}  
on the cosmological background shows that we just need to replace
several quantities in Eqs.~(\ref{hpf})-(\ref{Psip}) with 
$D \to d_{\rm GW}$, $f \to \tilde{f}=(1+z)^{-1}f$, 
$t_\infty \to t_{c}=(1+z) t_{\infty}$, 
and ${\cal M} \to \widetilde{\cal M}$, where 
$\widetilde{\cal M}$ is a chirp mass 
in the detector frame defined by \cite{Maggiore:2007ulw}
\be
\widetilde{\cal M}=(1+z) {\cal M}\,.
\ee
The same replacements can be also applied to the breathing 
scalar mode (\ref{hbfa}). 

Let us consider nonminially coupled theories given by the 
Horndeski functions (\ref{model1}), (\ref{model2}), 
and (\ref{model3}). 
In this case, we have  
$\kappa_0=[F/(2F+4\mu(\phi) X)]_{\phi=\phi_0}$ and 
$\zeta_0=[F+2\mu (\phi) X]|_{\phi=\phi_0}$.
We use an approximation that the background field 
value $\phi_0$ is constant in time and space, 
so that $X|_{\phi=\phi_0} \simeq 0$.
We also approximate $F(\phi_0) \simeq 1$ to recover 
the Einstein-Hilbert term $\Mpl^2R/2$ 
at large distances. Then, we have 
\be
\kappa_0 \simeq \frac{1}{2}\,,\qquad 
\zeta_0 \simeq 1\,.
\ee
The differences from $\kappa_0=1/2$ and $\zeta_0=1$  
work only as higher-order corrections to the 
scalar charge appearing in the phases and amplitudes 
of tensor and scalar GWs. 
Using the approximation $d_{\rm GW}(z) \simeq d_L (z)$, 
the resulting tensor and scalar gravitational 
waveforms are given by 
\ba
\tilde{h}_{+}(\tilde{f}) &=& -\tilde{h}_t (\tilde{f})(1+\cos^2 \iota)
e^{-i \Psi_{+}}\,,\label{hpFi} \\
\tilde{h}_{\times}(\tilde{f}) &=& -\tilde{h}_t (\tilde{f}) (2\cos \iota)
e^{-i \Psi_{\times}}\,,\\
\tilde{h}_{b}(\tilde{f}) &=& \tilde{h}_{s1}(\tilde{f}) (2\sin \iota) e^{-i\Psi_b}
+ \tilde{h}_{s2}(\tilde{f}) (2\sin^2 \iota) e^{-i\Psi_+}\,,
\label{hbFi}
\ea
where 
\ba
\tilde{h}_t (\tilde{f})
&=& \mathcal{A}_{\rm GR} (\tilde{f})
\left[ 1-\frac{5\eta^{2/5} \hat{\alpha}_A^2}{
96(G_* \widetilde{M} \pi \tilde{f})^{2/3}}
-\frac{\Gamma^2}{24}
\right]\,,\label{gwam}\\
\tilde{h}_{s1} (\tilde{f})
&=& \frac{\xi_0 \hat{\alpha}_A}{4}
\frac{\eta^{1/5}}{(G_* \widetilde{M} \pi \tilde{f})^{1/3}} \, 
\mathcal{A}_{\rm GR} (\tilde{f})
\,,\label{hfs1}\\
\tilde{h}_{s2} (\tilde{f})
&=& -\frac{\xi_0 \Gamma}{4}\, 
\mathcal{A}_{\rm GR} (\tilde{f})\,,\label{hfs2}\\
\Psi_{+}
&=& 
\Psi_{\times}+\frac{\pi}{2}
= 2\pi \tilde{f} t_{c}-2\Phi_c-\frac{\pi}{4}
+\frac{3}{128} \big(G_* \widetilde{\cal M} \pi 
\tilde{f}\big)^{-5/3}
\left[ 1-\frac{5\eta^{2/5} \hat{\alpha}_A^2}
{84 (G_* \widetilde{M} \pi \tilde{f})^{2/3}}
-\frac{\Gamma^2}{12} \right]\,,
\label{Psipl2}\\
\Psi_{b}
&=& 2\pi \tilde{f} t_{c}-\Phi_c-\frac{\pi}{4}
+\frac{3}{256 (2 G_* \widetilde{M} \pi \tilde{f})^{5/3}}
\left[ 1-\frac{5 \eta^{2/5} \hat{\alpha}_A^2}
{84 (2 G_* \widetilde{M} \pi \tilde{f})^{2/3}}
-\frac{\Gamma^2}{12} \right]\,,
\label{PsibF}
\ea
with
\be 
\mathcal{A}_{\rm GR} (\tilde{f})=
\sqrt{\frac{5\pi}{96}} \frac{(G_* \widetilde{\cal M})^{5/6}}
{d_L(z)} \left( \pi \tilde{f} \right)^{-7/6}\,.
\label{eq:GR-amplitude}
\ee
From Eqs.~(\ref{gwam}) and (\ref{hfs1}), the relative amplitude 
between the first scalar and tensor modes can be estimated as
$\tilde{h}_{s1} (\tilde{f})/\tilde{h}_{t} (\tilde{f}) \approx
\xi_0 \hat{\alpha}_A\,(c/v)$, where 
$v \approx (G_* \widetilde{M} \pi \tilde{f})^{1/3}$ is the 
relative circular velocity of the binary and we restored the speed of light $c$.
The other scalar-to-tensor ratio is of order 
$\tilde{h}_{s2} (\tilde{f})/\tilde{h}_{t} (\tilde{f})=
-\xi_0 \Gamma/4 \approx \xi_0 \hat{\alpha}_A$.
Provided that $\xi_0 \neq 0$ and $\hat{\alpha}_A \neq 0$, 
the breathing mode is nonvanishing relative to tensor polarizations.

In the waveforms (\ref{hpFi})-(\ref{hbFi}) with  
(\ref{gwam})-(\ref{PsibF}), there are two additional 
parameters $\hat{\alpha}_A$ and $\xi_0$ in comparison to those in GR. 
From the observations of the phases $\Psi_+$ and $\Psi_{\times}$ 
of tensor waves $\tilde{h}_{+}(\tilde{f})$ and 
$\tilde{h}_{\times}(\tilde{f})$, 
we expect that one can put constraints on the parameter $\hat{\alpha}_A$. 
The amplitude of the breathing scalar mode $\tilde{h}_{b}(\tilde{f})$ also allows 
a possibility of placing bounds on the product $\xi_0 \hat{\alpha}_A$. 
Thus, the observations of GWs emitted from the NS-BH inspiral binaries  
can provide constraints on both $\hat{\alpha}_A$ and $\xi_0$ simultaneously.

%%%%%%%%%%%%%%%%%%%%%%%%%%%%%%%%%%%%%%%%%%%%%%%%%%%%%%%%%%%%%%%%%%%%%%%%
\section{Parameterized framework for scalar-tensor GWs}
\label{sec:model}
%%%%%%%%%%%%%%%%%%%%%%%%%%%%%%%%%%%%%%%%%%%%%%%%%%%%%%%%%%%%%%%%%%%%%%%%

For the data analysis of scalar and tensor inspiral GWs from 
compact binary coalescences, we generalize the parameterized waveform 
model used in Ref.~\cite{Takeda:2021hgo}. 
In the following, we denote the chirp mass $\widetilde{\cal M}$ 
as ${\cal M}$ and the observed GW frequency $\tilde{f}$ 
as $f$ for simplicity.
We introduce a coupling parameter $\gamma$ between scalar modes and 
test masses of the detectors into the modified GW energy flux as 
\be
\label{eq:parameterized_flux}
\begin{split}
\dot{E}_{\rm GW} &= -\frac{d_{L}^2}{16\pi G_*}
\int \rd \Omega \left[ \langle \dot{h}_+^2 + 
\dot{h}_{\times}^2\rangle + \gamma \langle \dot{h}_b^2\rangle \right]\,,
\end{split}
\ee
where the angle bracket $\langle \cdots \rangle$ stands for 
an averaging procedure over the orbital evolution. 
In our theories, on using the relations $h_b=-\xi_0 \varphi/\Mpl$ and 
$\langle \partial_0 \varphi \partial_D \varphi \rangle
=-\langle \dot{\varphi}^2 \rangle$ and comparing Eq.~(\ref{eq:energy-loss}) with 
(\ref{eq:parameterized_flux}), the parameter $\gamma$ is given by 
\be
\gamma=\frac{1}{\kappa_0 \xi_0^2} 
\simeq \frac{2}{\xi_0^2}\,,
\ee
where we used $\kappa_0 \simeq 1/2$ 
in the second approximate equality.

For the time domain strain of each polarization with the modification of the gravitational constant $G_{\rm N}\rightarrow G_{*}$, which is degenerated with the change of intrinsic mass, we can assume the $\ell$-th harmonic of the orbital phase multiplied by the amplitude parameter $A_{p}^{(\ell)}$ and the inclination angle dependence $g^{(\ell)}_p(\iota)$~\cite{Takeda:2018uai} as
\begin{align}
\label{eq:pol_time}
h^{(\ell)}_p(t)=\frac{1}{2}A^{(\ell)}_p g^{(\ell)}_{p}(\iota) \frac{4 G_{*}{\cal M}}{d_L} \left( 2\pi G_{*} {\cal M} 
{\cal F} \right)^{2/3}
e^{-i\ell\Phi}\,,
\end{align}
where $p\in\{+, \times, b\}$ is a set of polarization indices 
running over plus ($+$), cross ($\times$), and breathing ($b$) modes. 
Here, ${\cal F}$ is the orbital frequency and $\Phi$ is the orbital phase. 
We consider only quadrupole radiation for the tensor polarizaiton. 
The non-zero amplitudes for the tensor modes are 
$A^{(2)}_{+}=A^{(2)}_{\times}=1$ by definition. 
On the other hand, we consider two types of radiation 
for the scalar modes: dipole $\ell=1$ and quadrupole $\ell=2$.

Applying the stationary phase approximation~\cite{Droz:1999qx, Maggiore:2007ulw, Yunes:2009yz} to Eq.~\eqref{eq:pol_time} and utilizing Eq.~\eqref{eq:parameterized_flux}, we can derive the 
frequency-domain GW signal in the form 
\be
\tilde{h}_I(f)=\tilde{h}_{T}(f)+\tilde{h}^{(1)}_{b}(f)
+\tilde{h}^{(2)}_{b}(f)\,,
\label{eq:signal_freq}
\ee
where $\tilde{h}_{T}(f)$, $\tilde{h}^{(1)}_{b}(f)$ 
and $\tilde{h}^{(2)}_{b}(f)$ represent the quadrupole tensor, 
dipole scalar, and quadrupole sclalar polarization 
contributions, respectively
\ba
\label{eq:hT_parameterized}
\tilde{h}_{T}(f) 
&=& -\left[F_I^{+}(1+\cos^2{\iota})-2iF_I^{\times}\cos{\iota}\right] 
\left[ 1+\delta A^{(2)} \right] 
\sqrt{\frac{5\pi}{96}}\frac{(G_* {{\cal M}})^2}{d_L}
\left( u_{*}^{(2)} \right)^{-7/2}
e^{-i\Psi_\text{GR}^{(2)}}e^{-i\delta\Psi^{(2)}}\,,\\
\label{eq:freq_b1_parameterized}
\tilde{h}^{(1)}_{b}(f)
&=& \sqrt{\frac{5\pi}{48}}A^{(1)}_{b}F_{I}^{b}(2\sin{\iota})
\eta^{1/5}\frac{(G_*{\cal M})^2}{d_L} 
\left( u_{*}^{(1)} \right)^{-9/2} 
e^{-i\Psi_\text{GR}^{(1)}}e^{-i\delta\Psi^{(1)}}\,,\\
\label{eq:freq_b2_parameterized}
\tilde{h}^{(2)}_{b}(f)
&=& \sqrt{\frac{5\pi}{96}}A^{(2)}_{b} 
F_{I}^{b}(2\sin^2{\iota})\frac{(G_*{\cal M})^2}{d_L}
\left( u_{*}^{(2)} \right)^{-7/2}e^{-i\Psi_\text{GR}^{(2)}}e^{-i\delta\Psi^{(2)}}\,.
\ea
Under the stationary phase approximation, the reduced $\ell$-th harmonic 
frequency is defined by 
\be
u^{(\ell)}_{*} \equiv \left( 
\frac{2\pi G_* {\cal M}f}{\ell} 
\right)^{1/3}\,,
\ee
where $f=\ell {\cal F}$ is the GW frequency. 
The modification of the gravitational constant $G_{*}/G_{\rm N}$ 
can be absorbed into the chirp mass like redshift. 
The functions $F_{I}^{A}\ (A=+,\times, b)$ are the antenna pattern functions 
of the $I$-th detector depending on the sky direction and the polarization 
angle, and representing the angular detector response to each polarization~\cite{Nishizawa:2009bf, Takeda:2020tjj}. $\Psi_{\rm GR}^{(\ell)}$ is the frequency evolution for the $\ell$-th harmonic in GR. Up to the second order 
with respect to $A^{(\ell)}_{b}$, 
the amplitude and phase corrections due to the backreaction of scalar radiation are given by
\begin{align}
\label{eq:amplitude_correction}
\delta A ^{(\ell)} &=
\delta A_{\rm d}^{(\ell)}+\delta A_{\rm q}^{(\ell)} \nonumber \\
&=-\frac{5}{48} \left(\tilde{A}_{b}^{(1)} \right)^2 
\eta^{2/5} \left(u_{*}^{(\ell)} \right)^{-2}
-\frac{1}{3} \left( \tilde{A}_{b}^{(2)} \right)^2\,, \\
\label{eq:phase_correction}
\delta \Psi^{(\ell)} &=
\delta \Psi_{\rm d}^{(\ell)}
+\delta \Psi_{\rm q}^{(\ell)} \nonumber \\
&=-\frac{5\ell}{3584} 
\left( \tilde{A}_{b}^{(1)} \right)^2
\eta^{2/5} \left( u_{*}^{(\ell)} \right)^{-7}
-\frac{\ell}{128} \left( \tilde{A}_{b}^{(2)} \right)^2 
\left( u_{*}^{(\ell)} \right)^{-5} \;,
\end{align}
where  
\be
\tilde{A}^{(\ell)}_{b}=\sqrt{\gamma} A^{(\ell)}_{b}\,.
\label{tA}
\ee

In comparison to GR, the above waveform contains four additional 
parameters: two scalar GW amplitude parameters $A^{(1)}_{b}$ and $A^{(2)}_{b}$, 
and two phase evolution parameters $\tilde{A}^{(1)}_{b}$ and 
$\tilde{A}^{(2)}_{b}$. 
In the original work~\cite{Takeda:2021hgo}, it is assumed that the 
stress-energy tensor is the same form as in GR, that is, $\gamma=1$ and then $\tilde{A}^{(\ell)}_{b}$ is identical to $A^{(\ell)}_{b}$. 
In our scalar-tensor theories, the quantity $\gamma$ in Eq.~(\ref{tA})  
is given by $\gamma=2/\xi_0^2$, where $\xi_0=\Mpl F_{,\phi}/F$. 
For the nonminimal couplings $F=e^{-2Q\phi/\Mpl}$ and 
$F=e^{-\beta \phi^2/(2\Mpl^2)}$, we have $\xi_0=-2Q$ and 
$\xi_0=-\beta \phi/\Mpl$, respectively. 
Thus, the phase evolution parameters 
$\tilde{A}^{(\ell)}_{b}=\sqrt{2/\xi_0^2} A^{(\ell)}_{b}$ 
keep not only information on the amplitudes but also that 
on the nonminimal couplings. 
The above waveform model matches the generalized parameterized post-Einsteinian framework~\cite{Chatziioannou:2012rf, Tahura:2018zuq, Zhang:2019iim} limited to the scalar-tensor polarizations. 
The analysis based on this waveform model provides a very general 
framework to search for GW polarizations from compact binary coalescences 
in massless scalar-tensor theories.

Now, we compare our theoretical waveforms given in Eqs.~(\ref{hpFi})-(\ref{PsibF}) 
with Eqs.~\eqref{eq:hT_parameterized}-\eqref{eq:freq_b2_parameterized}. 
Then, we obtain the following correspondences:
\begin{align}
\label{eq:parms_relations}
A^{(1)}_{b} &=\frac{1}{2}\xi_0 \hat{\alpha}_A \;, 
\qquad A^{(2)}_{b}= -\frac{1}{4}\Gamma \xi_0 =\frac{2\eta}{1-\sqrt{1-4\eta}}A^{(1)}_b  \;, \qquad  \\
\label{eq:parms_relations2}
\tilde{A}^{(1)}_{b} &= \frac{1}{\sqrt{2}} \hat{\alpha}_A \;, \qquad 
\tilde{A}^{(2)}_{b} =-\frac{\sqrt{2}}{4}\Gamma  
= \frac{2\eta}{1-\sqrt{1-4\eta}}\tilde{A}^{(1)}_{b}\,,
\end{align}
where we assumed that $m_A<m_B$.
Thus, it is enough to consider only two additional deviation parameters $A^{(1)}_{b}$ and $\tilde{A}^{(1)}_{b}$, which depend on the two 
physical quantities $\hat{\alpha}_A$ and $\xi_0$.

%%%%%%%%%%%%%%%%%%%%%%%%%%%%%%%%%%%%%%%%%%%%%
\section{Observational constraints from GW200115}
\label{sec:data_analysis}
%%%%%%%%%%%%%%%%%%%%%%%%%%%%%%%%%%%%%%%%%%%

As described in Section~\ref{sec:model}, not only the non-tensorial polarization modes are observables that exhibit a signature of GR violation, but also the constraints on them can bring us independent information on the possibility of extension of GR apart from the modification of the tensor phase evolution. In addition, the lack of non-tensorial polarizations in GW analysis may cause parameter bias and over/underestimate the constraints on the specific theoretical parameters. 
It is worth mentioning that techniques like null streams, as performed in \cite{LIGOScientific:2018czr,LIGOScientific:2021sio,Hagihara:2019ihn}, provide model-independent methods for detecting signs of non-tensorial polarization. These methods, while not yielding direct parameter constraints, 
contribute to the weaker but model-agnostic search 
for beyond-GR polarizations.

In this section, we analyze the GW signal of the compact binary merger event listed in LIGO-Virgo-KAGRA catalog under the scalar-tensor polarization waveform model~\eqref{eq:signal_freq} to provide constraints on the maximal observables that can be extracted from full GW polarizations. In Section~\ref{sec:theory}, we demonstrate that one can obtain constraints on the multiple theoretical parameters from a single GW observation. In particular, we put constraints on both of the scalar charge of compact stars and the coupling of the scalar polarization with matter. 

\subsection{Data analysis}
\label{sec:Data_analysis}

We focus on the NS-BH merger where only the NS has a scalar charge in the 
framework of luminal Horndeski theories. Separation of polarization modes by the current GW detector network requires at least as same operating detectors as the polarization modes~\cite{Takeda:2018uai}. We can safely select GW200115 for analysis because it is a only plausible binary NS-BH coalescence event observed by three detector network so far~\cite{LIGOScientific:2021qlt}. The primary mass is within the mass range of known black holes and the secondary mass is within that of known neutron stars. 
We use the strain data of GW200115 from the Gravitational Wave Open Science Center~\cite{LIGOScientific:2019lzm, KAGRA:2023pio}, which is down sampled to $2048\ {\rm Hz}$ to reduce computational cost.
The signal duration considered for parameter estimation is $64\ {\rm sec}$, with a post-merger duration of $2\ {\rm sec}$.

We analyze the signal of GW200115 in the scalar-tensor polarization framework 
where the GW signal is described by Eq.~\eqref{eq:signal_freq}. 
There are two additional parameters $A^{(1)}_b$ and $\tilde{A}^{(1)}_b$
in addition to standard 11 source parameters: the primary and secondary masses in the detector frame, $m_A$ and $m_B$, the dimensionless spins for aligned spin binaries, $\chi_A$ and $\chi_B$, the luminosity distance to the compact binary system $d_L$, the inclination angle $\iota$, the right ascension and declination of the compact binary system, $\alpha$ and $\delta$, polarization angle $\psi$, the coalescence time $t_c$, and the phase at the reference frequency $\phi_{\rm ref}$. We do not include a tidal effect on the secondary because the tidal effects are rarely effective~\cite{LIGOScientific:2021qlt}. Hence, we fix the tidal deformability parameters $\Lambda_A$ and $\Lambda_B$ to zero. In the analysis, we use the flat $\Lambda$ cold dark matter cosmological model whose parameters are given by the results of Planck13~\cite{Planck:2013pxb}. 
Note that using the Planck18 data practically give 
the same results as those presented below.
Hence, a set of the parameters $\bm{\theta}$ is given by 
\begin{align}
\bm{\theta} \equiv 
\left\{ m_A, m_B, \chi_A, \chi_B, d_L,\iota, \alpha, \delta, \psi, t_c, \phi_{\rm ref}, A^{(1)}_{b},
\tilde{A}^{(1)}_{b} 
\right\}\,.
\end{align}

Our analysis relies on the Bayesian inference. 
Given a hypothetical model $M$ described by a set of parameters $\bm{\theta}$ and 
a set of detector signals $\bm{d}$, the posterior probability 
distribution $p(\bm{\theta}| \bm{d}, M)$ is computed through the Bayes' 
theorem as~\cite{Maggiore:2007ulw, LIGOScientific:2019hgc}
\be
p(\bm{\theta}| \bm{d}, M) = 
\frac{p(\bm{\theta}| M) p(\bm{d} | \bm{\theta}, M)}
{p(\bm{d}| M)}\,,
\ee
where $p(\bm{\theta}| M)$ is the prior probability distribution, 
and $p(\bm{d} | \bm{\theta}, M)$ is the likelihood. Under the assumptions 
that the detector noise is stationary and Gaussian, we use the standard 
Gaussian likelihood,
\begin{align}
p(\bm{d} | \bm{\theta}, M)\propto \exp{\left[-\frac{1}{2} 
\sum_{I} \langle h_{I}(\bm{\theta})-d_I | h_{I}(\bm{\theta})-d_I \rangle \right]}\,,
\end{align}
where $I$ is the detector label. The angle bracket 
$\langle\ |\ \rangle$ represents the noise-weighted 
inner product defined by
\begin{align}
\langle a | b \rangle \equiv 
4\,{\rm Re} \int_{f_{\rm min}}^{f_{\rm max}}
\frac{a^*(f) b(f)}{S_{I, n}(f)} \rd f\,,
\end{align}
where $S_{I, n}(f)$ is the noise power spectral density of the $I$-th detector. Here, the lower cutoff frequency $f_{\rm min}$ is set to $20\ {\rm Hz}$ for LIGO Hanford and Vigro detectors, but $25\ {\rm 
Hz}$ for LIGO Livingston detector due to the scattering noise around $20\ {\rm Hz}$~\cite{LIGOScientific:2021qlt}. 
On the other hand, the upper cutoff frequency $f_{\rm max}$ is set to the inner most stable circular orbit frequency for nonspinning objects in GR\footnote{The upper cutoff frequency $f_{\rm ISCO}$ is also modified in the scalar-tensor theory and can be different from the GR value used in our data analysis. However, as $f_{\rm ISCO}$ in our case is determined mostly by the non-scalarized BH mass $m_B$ and the modification 
on $m_A$ can change $f_{\rm ISCO}$ only by several percents at most, we neglect the effect here.}, $f_{\rm ISCO}= (6^{3/2}\pi m)^{-1}\simeq 604\ {\rm Hz}$, to restrict our analysis in the binary inspiral stage. 
Instead of estimating the noise power spectral density from strain data by the Welch method, we use the event specific power spectral density available in LVK posterior sample releases~\cite{Niu:2021nic}. For the Bayesian inference, we utilize the \texttt{Bilby} software~\cite{Ashton:2018jfp} and the \texttt{dynesty} sampler~\cite{Speagle:2019ivv}. The sampler settings are chosen by referring to~\cite{Romero-Shaw:2020owr}. Specifically, we set the number of live points (nlive) to 2048 and adopted an acceptance-walk strategy with 200 walks. The action number (nact) was set to 10. For all analysis in this paper, we confirmed that the results do not change significantly with increased number of live points.

As an inspiral template in GR, we adopt \texttt{IMRPhenomD\_NRTidal}~\cite{Dietrich:2019kaq} implemented in the LIGO Algorithm Library \texttt{LALSuite}~\cite{lalsuite} in which the inspiral GW phase of the $\ell$-th harmonic is given by
\begin{align}
\Psi^{(\ell)}(f)=2\pi f t_c -\ell \phi_{\rm ref} -\frac{\pi}{4} + \Psi^{(\ell)}_{\rm PN\ series}(f)
-\Psi^{(\ell)}_{\rm PN\ series}(f_{\rm ref})\,,
\end{align}
with the post-Newtonian expansion series
\begin{align}
\Psi^{(\ell)}_{\rm PN\ series}(f)=
\frac{3 \ell}{256} \left( u_{*}^{(\ell)} \right)^{-5}
\sum_{i=0}^{7}\phi_i \left( u_{*}^{(\ell)} \right)^i\,,
\label{eq:PN_expansion}
\end{align}
where $\phi_i$ is the PN coefficients compiled in~\cite{Khan:2015jqa}. The reference frequency is set to be $20\ {\rm Hz}$ in this work. 
We note that the \texttt{IMRPhenomD\_NRTidal} modelling 
is based on an inspiral-merger-ringdown waveform, whose 
late-inspiral phase evolution has been calibrated by 
the numerical-relativity waveform.
Also note that \texttt{IMRPhenomD\_NRTidal} consists of only 
the leading-order quadrupolar mode.

For the priors, we use the same priors used in LVK analysis of GW200115 for the standard binary parameters~\cite{LIGOScientific:2021qlt}. As for the spin of 
the NS, we use a low spin prior from the NS property. 
In order to make it possible to reveal the ability of the GW observations to put constraints on the scalar-tensor theories, and interpret the results even in the other theories predicting scalar polarization, we apply uniform priors in the range $[-1, 1]$ for $A^{(1)}_{b}$ and $\tilde{A}^{(1)}_{b}$ without invoking the constraints from the solar system experiments and the binary pulsar observations.

\subsection{Results}
\label{sec:results}

\subsubsection{Scalar-tensor theories}
We performed the Bayesian analysis for GW200115 signal with the scalar-tensor polarization waveform model Eq.~\eqref{eq:signal_freq} following the prescription described in Section \ref{sec:Data_analysis}. 

\begin{figure}[h]
\begin{center}
\includegraphics[height=6.5in,width=6.8in]{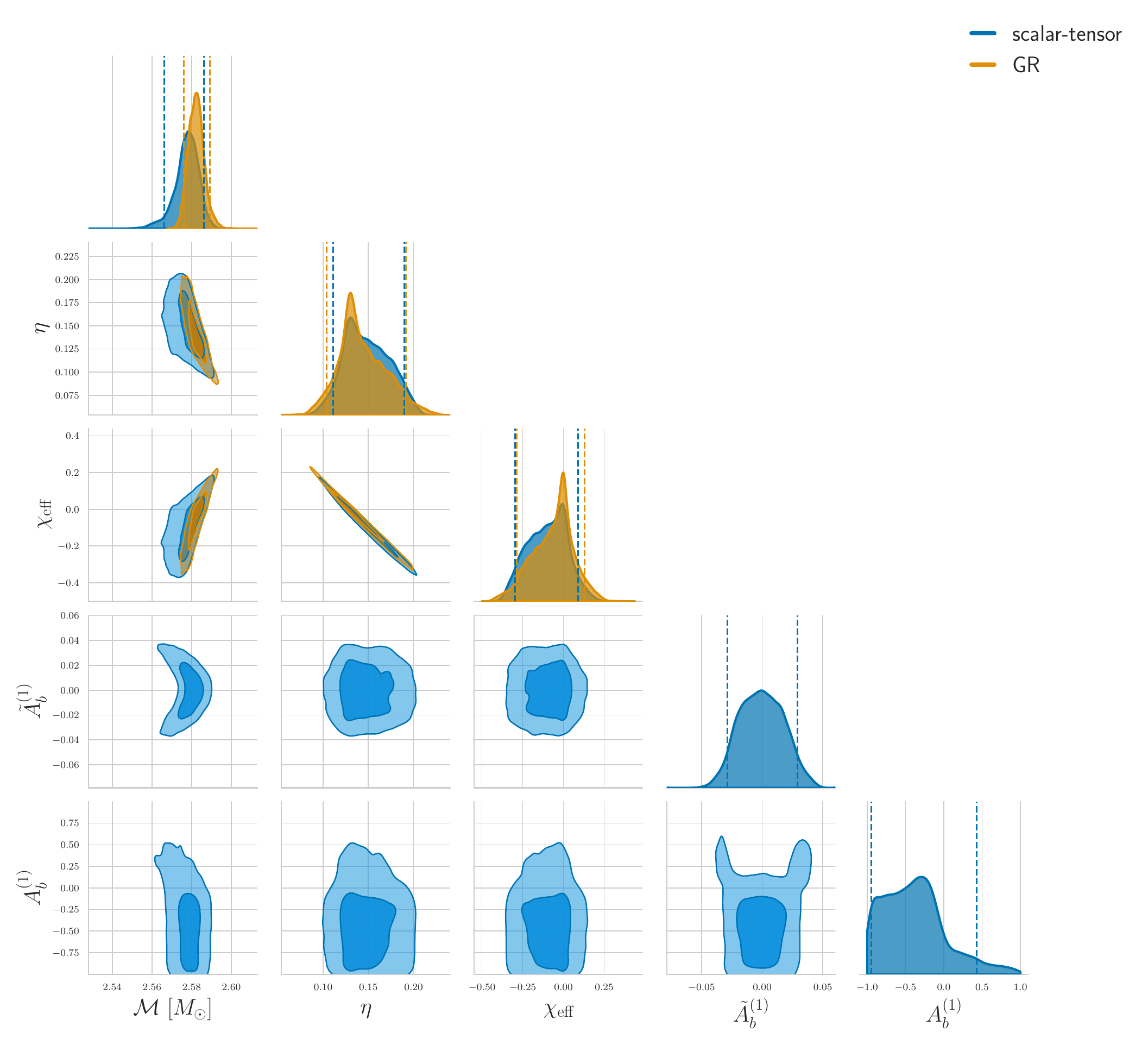}
%comparison_posterior_plot_Mc_eta_chieff_tildeAb1_Ab1_seaborn.py
\end{center}
\caption{The posterior probability distributions of the phase parameters and the scalar radiation parameters, the chirp mass in the detector frame $\mathcal{M}$, the luminosity distance $d_L$, the inclination angle $\iota$, and the additional parameters due to scalar radiation $\tilde{A}^{(1)}_{b}$ and $A_{b}^{(1)}$, are shown in blue for GW200115 with 50\% and 90\% credible intervals 
in the 2-dimensional plots and 90\% credible interval in the 1-dimensional plots. 
For comparison, the results under GR by LVK collaboration~\cite{LIGOScientific:2021qlt} 
are also shown in orange. The estimated mean values 
is $\tilde{A}^{(1)}_{b}=-0.001^{+0.030}_{-0.028}$ with 90\% credible level.}
\label{fig:corner_phase_widest}
\end{figure}

Fig.~\ref{fig:corner_phase_widest} shows the posterior probability distribution 
of the phase parameters such as chirp mass in the detector frame $\mathcal{M}$, 
symmetric mass ratio $\eta$, effective spin $\chi_{\rm eff}$, 
and the additional deviation parameters ${A}^{(1)}_{b}$ and 
$\tilde{A}^{(1)}_{b}$ under the scalar-tensor model in blue. In corner plots, we draw 50\% and 90\% credible intervals in the 2-dimensional plots and 90\% credible intervals in the 1-dimensional plots. For comparison, we also show the posterior distribution estimated from the GR analysis by LVK collaboration~\cite{LIGOScientific:2021qlt} in orange. The additional phase parameter $\tilde{A}^{(1)}_{b}$ is strongly correlated with the chirp mass $\mathcal{M}$, while we did not find any strong correlation of $\tilde{A}^{(1)}_{b}$ with the other phase parameters clearly.  As shown in Fig.~\ref{fig:corner_phase_widest}, we also find that there are no strong correlation between ${A}^{(1)}_{b}$ and the phase parameters. This is reasonable because ${A}^{(1)}_{b}$ is the parameter characterizing the scalar GW amplitudes. As a result, we obtain the bound with 90\% credible level,
\be
\tilde{A}^{(1)}_{b}=-0.001^{+0.030}_{-0.028}\,.
\label{eq:constraint_tilde_A}
\ee

\begin{figure}[h]
\begin{center}
\includegraphics[height=6.5in,width=6.8in]{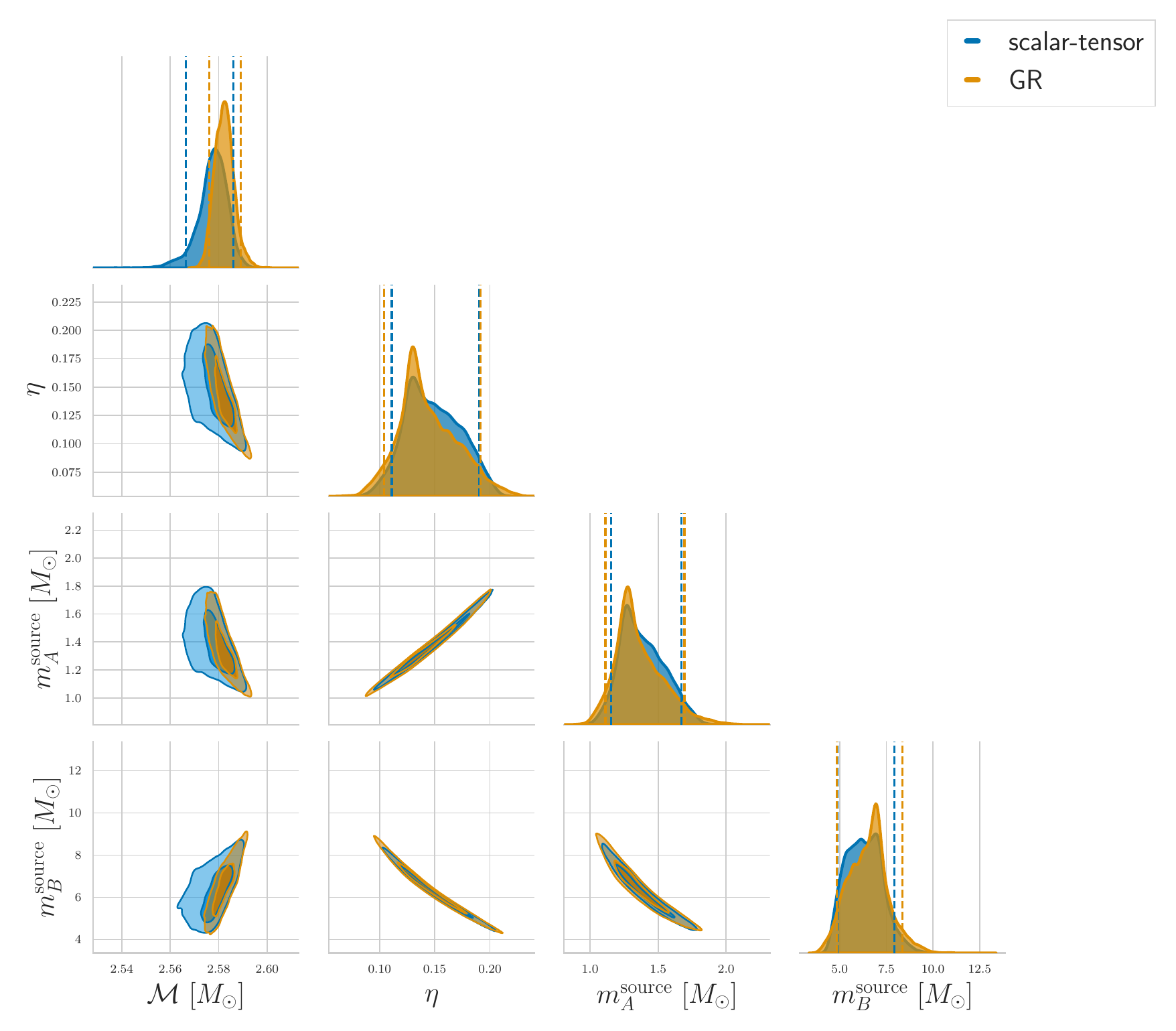}
%comparison_posterior_plot_Mc_dL_iota_tildeAb1_Ab1_seaborn.py
\end{center}
\caption{The posterior probability distributions of the mass parameters, the chirp mass in the detector frame $\mathcal{M}$, the symmetric mass ratio $\eta$, and the component masses in the source frame $m_A^{\rm source}$ and $m_B^{\rm source}$, are shown in blue for GW200115 with 50\% and 90\% credible intervals 
in the 2-dimensional plots and 90\% credible interval in the 1-dimensional plots. 
For comparison, the results under GR by LVK collaboration~\cite{LIGOScientific:2021qlt} 
are also shown in orange.}
\label{fig:mass_parameters_widest}
\end{figure}

The correlation between $\tilde{A}^{(1)}_{b}$ and $\mathcal{M}$ would come from the phase dependence in the tensor phase evolution like
$(u_{*}^{(\ell)})^{-5}
[1-(5/84)\eta^{2/5}\tilde{A}_{b}^{(1)2} (u_{*}^{(\ell)})^{-2}]$,
in which the first term comes from the GR contribution at 0PN in Eq.~\eqref{eq:PN_expansion} and the second term originates from the first term corresponding to the scalar dipole radiation in Eq.~\eqref{eq:phase_correction},
to compensate each other such that the overall phase 
is kept to the constant. Due to the correlation, the posterior distribution of the chirp mass is 
biased to smaller value compared to that under GR. Fig.~\ref{fig:mass_parameters_widest} shows the distributions for mass parameters. Reflecting the bias in the estimation of the chirp mass, the estimation of the component masses are slightly affected with respect to the GR case, but the value still infers that the smaller compact star is a NS, which is constrained as
\begin{align}
\label{eq:component_mass_constraint}
    1.15M_{\odot}<m_A<1.67M_{\odot}\,,
\end{align}
with 90\% credible level. 

%%%%%%%%%%%%%%%%%%%%%%%%%%%%%%%%%%%%%%%%%%%%%%%%%%%%%%%%%%%%%%%%%%%%%%%%%%
%%%%%%%%%%%%%%%%%%%%%%%%%%%%%%%%%%%%%%%%%%%%%%%%%%%%%%%%%%%%%%%%%%%%%%%%%%%

%%%%%%%%%%%%%%%%%%%%%%%%%%%%%%%%%%%%%%%%%%%%%%%%%%%%%%%%%%%%%%%%%%%%%%%%%%
%%%%%%%%%%%%%%%%%%%%%%%%%%%%%%%%%%%%%%%%%%%%%%%%%%%%%%%%%%%%%%%%%%%%%%%%%%%

\begin{figure}[ht]
\begin{center}
\includegraphics[height=6.5in,width=6.8in]{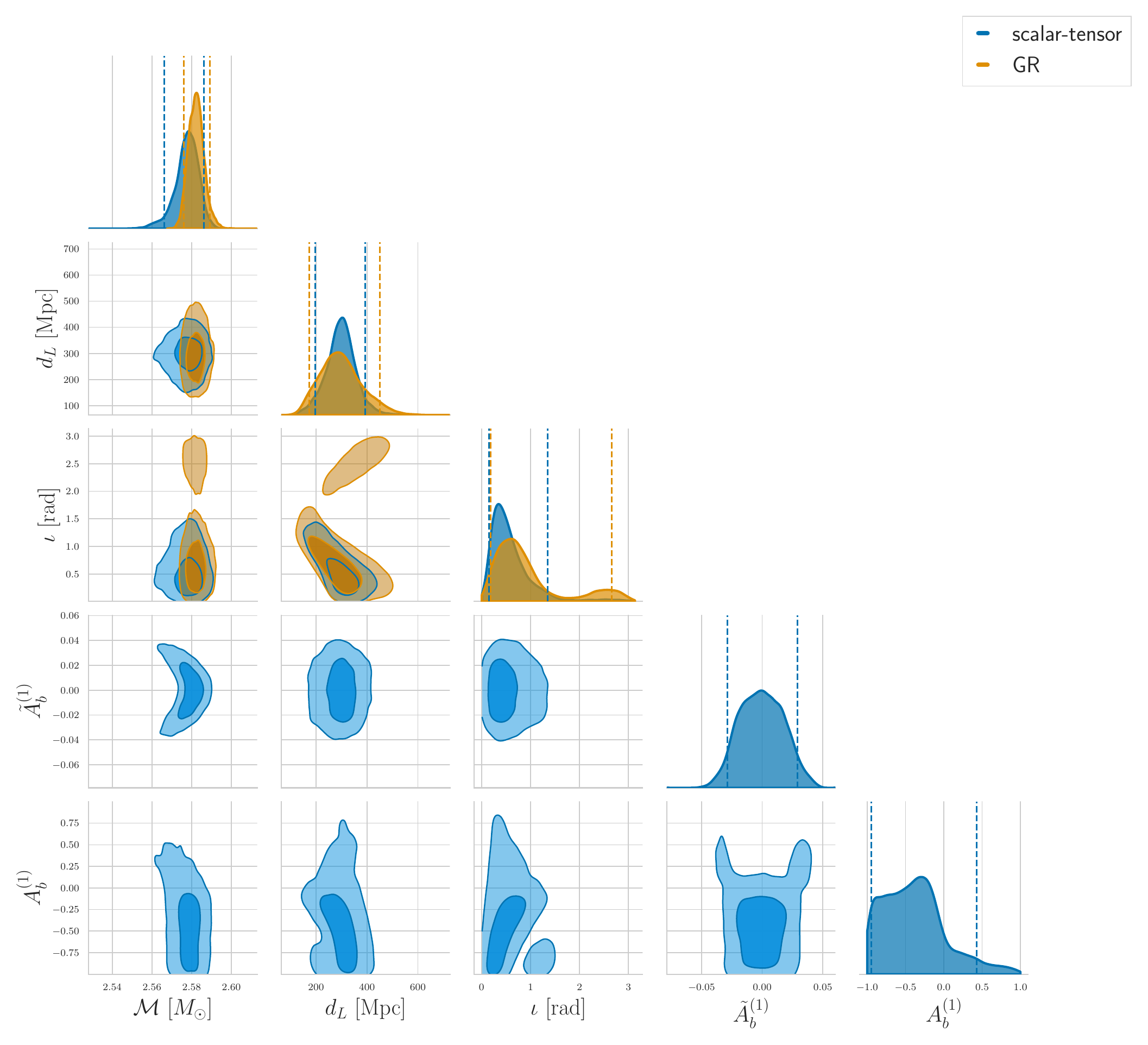}
\end{center}
\caption{The posterior probability distributions of the amplitude parameters and 
the scalar polarization parameters: the chirp mass in the detector frame $\mathcal{M}$, 
the luminosity distance $d_L$, the inclination angle $\iota$, and the additional 
parameters due to scalar radiation $\tilde{A}_{b}^{(1)}$ and $A_{b}^{(1)}$, for GW200115 
with 50\% and 90\% credible intervals in the 2D plots and 90\% credible interval in the 1D plots. 
For comparison, the results under GR by LVK collaboration~\cite{LIGOScientific:2021qlt} 
are also shown in orange. We found little correlations with the amplitude parameters, 
the luminosity distance $d_{L}$ and the inclination angle $\iota$, and 
the scalar-mode amplitude parameter $\tilde{A}_{b}^{(1)}$. The estimated mean value is $A_{b}^{(1)}=-0.41^{+0.84}_{-0.54}$ with 90\% credible level. The constraint on the scalar GW amplitude $A^{(1)}_b$ 
can be converted into the constraint on the scalar-to-tensor amplitude ratio defined by Eq.~\eqref{eq:def_scalar_to_tensor_ratio}: $R^{(1)}_{ST} \leq 0.57$. 
}
\label{fig:corner_amplitude_widest}
\end{figure}

Fig.~\ref{fig:corner_amplitude_widest} shows the posterior probability distribution 
of the amplitude parameters such as chirp mass in the detector frame $\mathcal{M}$, luminosity distance $d_L$, inclination angle $\iota$, and the additional scalar  parameters $A^{(1)}_{b}$ and $\tilde{A}^{(1)}_{b}$ in blue.
In corner plots, we draw 50\% and 90\% credible intervals in the 2-dimensional plots and 90\% credible intervals in the 1-dimensional plots again. For comparison, we also show the posterior distribution estimated from the GR analysis by LVK collaboration~\cite{LIGOScientific:2021qlt} in orange.
We do not find correlations between the phase parameter $\tilde{A}^{(1)}_{b}$, and the luminosity distance $d_L$ or the inclination angle $\iota$.
However, there is little correlation of the scalar GW amplitude $A^{(1)}_{b}$ with the other amplitude parameters. 
Note that the inclusion of the scalar polarization seems to improve the inclination determination. We will discuss this point in detail at the end of this section by comparing the results based on different analytical settings.
With the current detector sensitivity, it was found that some scalar GW amplitude samples hit the edge of the prior. However, since the 90\% CL is inside the prior, we found the first constraint on the additional 
parameter purely characterizing scalar polarization amplitude
\be
A^{(1)}_{b}=-0.41^{+0.84}_{-0.54}\,,
\label{eq:constraint_A}
\ee
with 90\% credible interval. 

%%%%%%%%%%%%%%%%%%%%%%%%%%%%%%%%%%%%%%%%%%%%%%%%%%%%%%%%%%%%%%%%%%%%%%%%%%
%%%%%%%%%%%%%%%%%%%%%%%%%%%%%%%%%%%%%%%%%%%%%%%%%%%%%%%%%%%%%%%%%%%%%%%%%%

\begin{figure}[ht]
\begin{center}
\includegraphics[height=3.8in,width=4.5in]{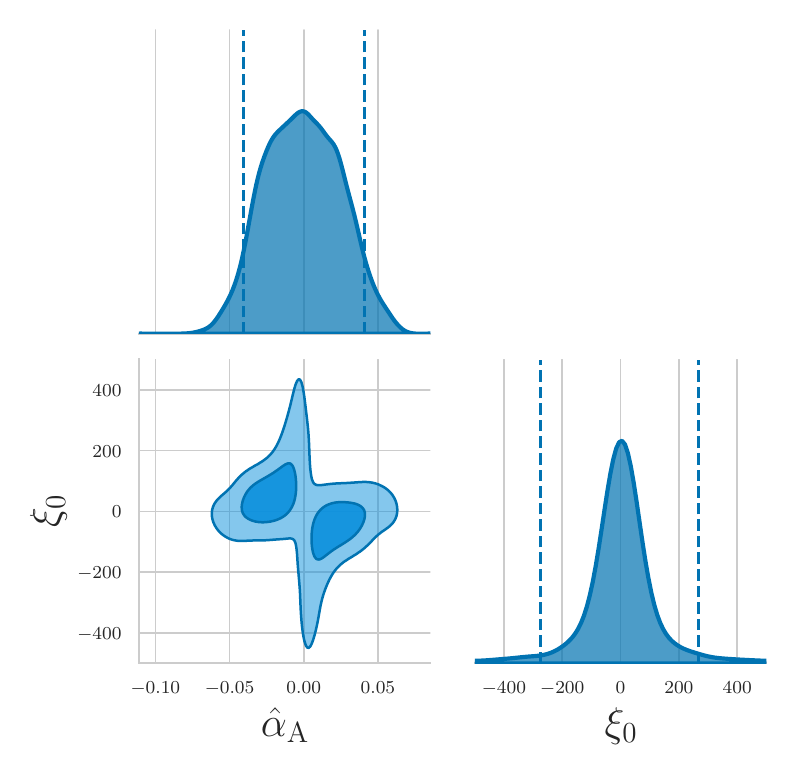}
\end{center}
\caption{
The posterior probability distributions of the additional phase correction $\tilde{A}_{b}^{(1)}$ 
and the scalar GW amplitudes $A_{b}^{(1)}$ for GW200115 with 50\% and 90\% credible intervals 
in the 2-dimensional plots and 90\% credible interval in the 1-dimensional plots. 
The constraints on $\tilde{A}^{(1)}_{b}=-0.001^{+0.030}_{-0.028}$ and 
$A_{b}^{(1)}=-0.41^{+0.84}_{-0.54}$ when considering quadrupole scalar radiation 
can be converted into the bounds on the Horndeski parameters: 
$\hat{\alpha}_A=-0.001^{+0.042}_{-0.040}$ and 
$\xi_{0}=3.10^{+265}_{-279}$ with 90\% credible intervals. The sparse samples of $\hat{\alpha}_A\sim 0$ in the 2D plot is due to the fact that $\xi_0$ is so large for the samples around $\tilde{A}^{(1)}_{b}\sim 0$, where some samples are sparsely distributed outside the depicted region of the figure.}
\label{fig:corner_alpha_xi_widest2}
\end{figure}

On using the correspondence (\ref{eq:parms_relations2}) with the constraint on $\tilde{A}^{(1)}_b$ Eq.~\eqref{eq:constraint_tilde_A}, 
the amplitude of $\hat{\alpha}_A$ is constrained to be 
\be
|\hat{\alpha}_A| \leq 0.041\,,
\ee
at 90\% credible level. This gives the upper limit on the amount of the NS scalar charge in the mass range Eq.~\eqref{eq:component_mass_constraint}.
The vanishing scalar charge ($\hat{\alpha}_A=0$) is consistent 
with the GW200115 data.  
On the other hand, on using the correspondence (\ref{eq:parms_relations}) on ${A}^{(1)}_b$ Eq.~\eqref{eq:constraint_A}, the product 
$\xi_0 \hat{\alpha}_A$ is constrained to be 
\be
-1.88<\xi_0 \hat{\alpha}_A<0.86 \,.
\ee
Compared to $\hat{\alpha}_A$, we only have a weak bound on 
the other parameter $\xi_0$.
This is associated to the fact that the amplitude of 
the breathing polarization is poorly constrained 
with the GW200115 data.
Fig.~\ref{fig:corner_alpha_xi_widest2} shows the posterior distributions on two 
theoretical model parameters, the scalar charge $\hat{\alpha}_A$ and 
the nonminimal coupling strength converted from the samples of 
$\tilde{A}_{b}^{(1)}$ and $A_{b}^{(1)}$ through the relations \eqref{eq:parms_relations}. 
We put constraints on these scalar radiation parameters simultaneously only from the observation of the single compact binary coalsecence event. 
The estimated values are 
\ba 
\hat{\alpha}_A &=& 
-0.001^{+0.042}_{-0.040}\,,\label{aAbound} \\
\xi_{0} &=& 3.10^{+265}_{-279}\,,\label{xi0bound}
\ea
with 90\% credible intervals.

%%%%%%%%%%%%%%%%%%%%%%%%%%%%%%%%%%%%%%%%%%%%%%%%%%%%%%%%%%%%%%%%%%%%%%%%%%
%%%%%%%%%%%%%%%%%%%%%%%%%%%%%%%%%%%%%%%%%%%%%%%%%%%%%%%%%%%%%%%%%%%%%%%%%%

\subsubsection{Comparison with different waveform components}
In order to assess the contribution of the amplitude and phase parts of each polarization mode to the posterior samples, we analyze GW200115 with two different settings of the waveform model: (i) {\it tensor modes with scalar corrections} and (ii) {\it tensor modes plus scalar dipole mode}.\\

\begin{figure}[h]
\begin{center}
\includegraphics[height=6.5in,width=6.8in]{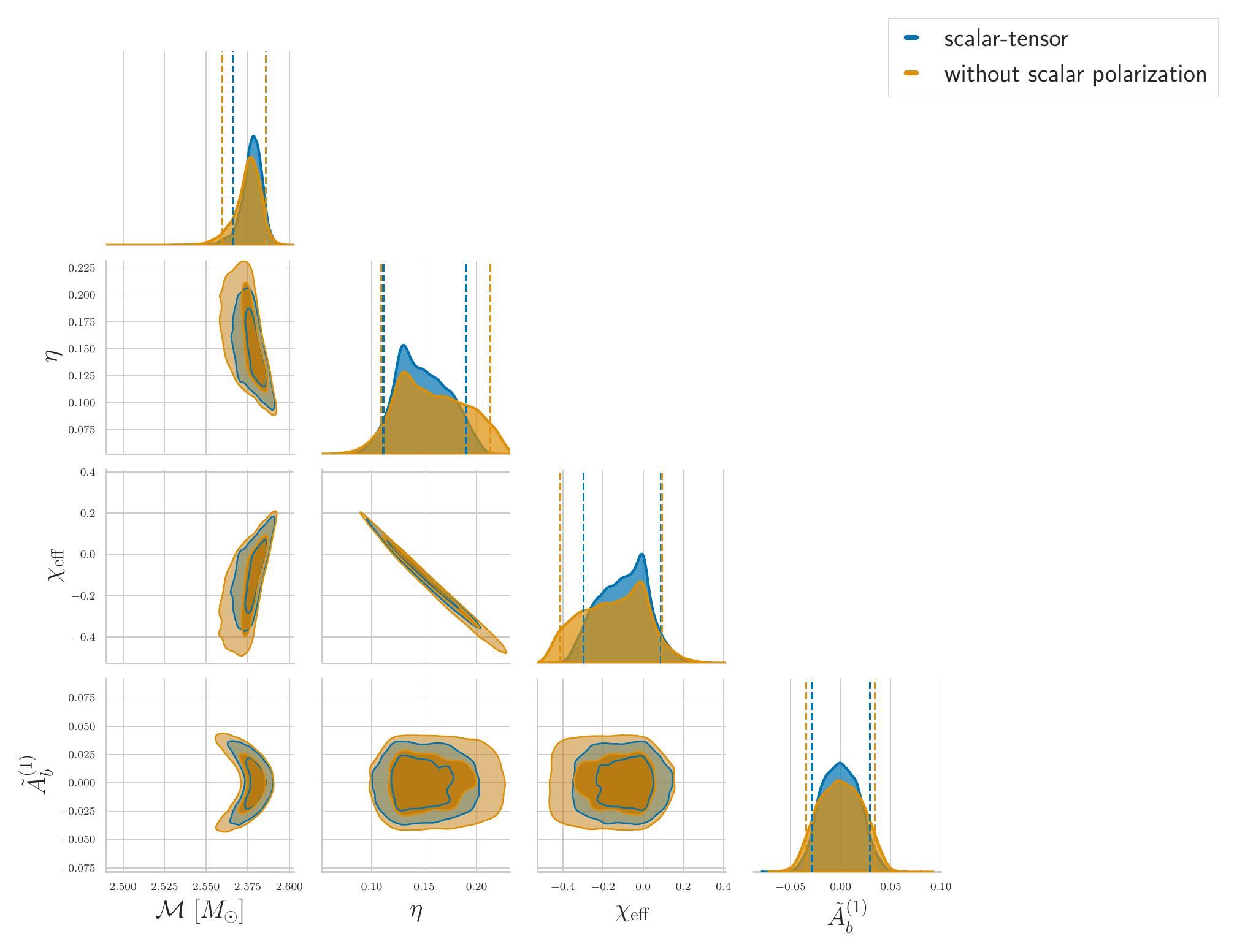}
\end{center}
\caption{This is a similar figure to Fig.~\ref{fig:corner_phase_widest} but compared to the analysis using the model of tensor modes with scalar phase corrections.
The estimated mean values 
is $|\tilde{A}^{(1)}_{b}|\leq 0.034$ with 90\% credible level in comparison with $|\tilde{A}^{(1)}_{b}|\leq 0.029$ for the scalar-tensor model.}
\label{fig:corner_phase_comparison_without_scalar}
\end{figure}

\paragraph{Tensor modes with scalar phase corrections}
First, we perform the Bayesian analysis based on the waveform model including only the tensor modes $\tilde{h}_{+,\times}$ with the amplitude and phase corrections due to scalar radiation. In this model, we do not consider the appearance of scalar polarization modes, Eqs.~\eqref{eq:freq_b1_parameterized} and~\eqref{eq:freq_b2_parameterized}. Hence, the waveform model is given by
\be
\tilde{h}_I(f)=\tilde{h}_{T}(f)\,,
\ee
where the GW polarizations are
\ba
\tilde{h}_{T}(f) 
&=& -\left[F_I^{+}(1+\cos^2{\iota})-2iF_I^{\times}\cos{\iota}\right]  
\sqrt{\frac{5\pi}{96}}\frac{(G_* {{\cal M}})^2}{d_L}
\left( u_{*}^{(2)} \right)^{-7/2}
e^{-i\Psi_\text{GR}^{(2)}}e^{-i\delta\Psi^{(2)}}\,,
\ea
with the phase corrections
\begin{align}
\delta \Psi^{(\ell)} &=-\frac{5\ell}{3584} 
\left( \tilde{A}_{b}^{(1)} \right)^2
\eta^{2/5} \left( u_{*}^{(\ell)} \right)^{-7}
-\frac{\ell}{128} \left( \tilde{A}_{b}^{(2)} \right)^2 
\left( u_{*}^{(\ell)} \right)^{-5} \;.
\end{align}
This reduced waveform model imitates the parameterized tests for the inspiral GWs performed by LVK collaboration~\cite{LIGOScientific:2021sio}. The difference between this reduced analysis and the parameterized tests by the LVK collaboration is essentially the prior setting for the phase deviation parameter. In our model, we adopt a uniform prior for $\tilde{A}^{(1)}_{b}$, but they use a uniform prior on $\varphi_{-2}$, which is corresponding to $\sim (\tilde{A}^{(1)}_{b})^2$.
Hence, we can evaluate the contribution of the existence of scalar polarization itself to the parameter estimation by comparing this analysis with the results of the scalar-tensor model. Fig.~\ref{fig:corner_phase_comparison_without_scalar} shows the comparison of the posterior samples for the phase parameters between the scalar-tensor model (blue) and the model of tensor modes with scalar corrections (orange). 

\begin{figure}[ht]
\begin{center}
\includegraphics[height=6.5in,width=6.8in]{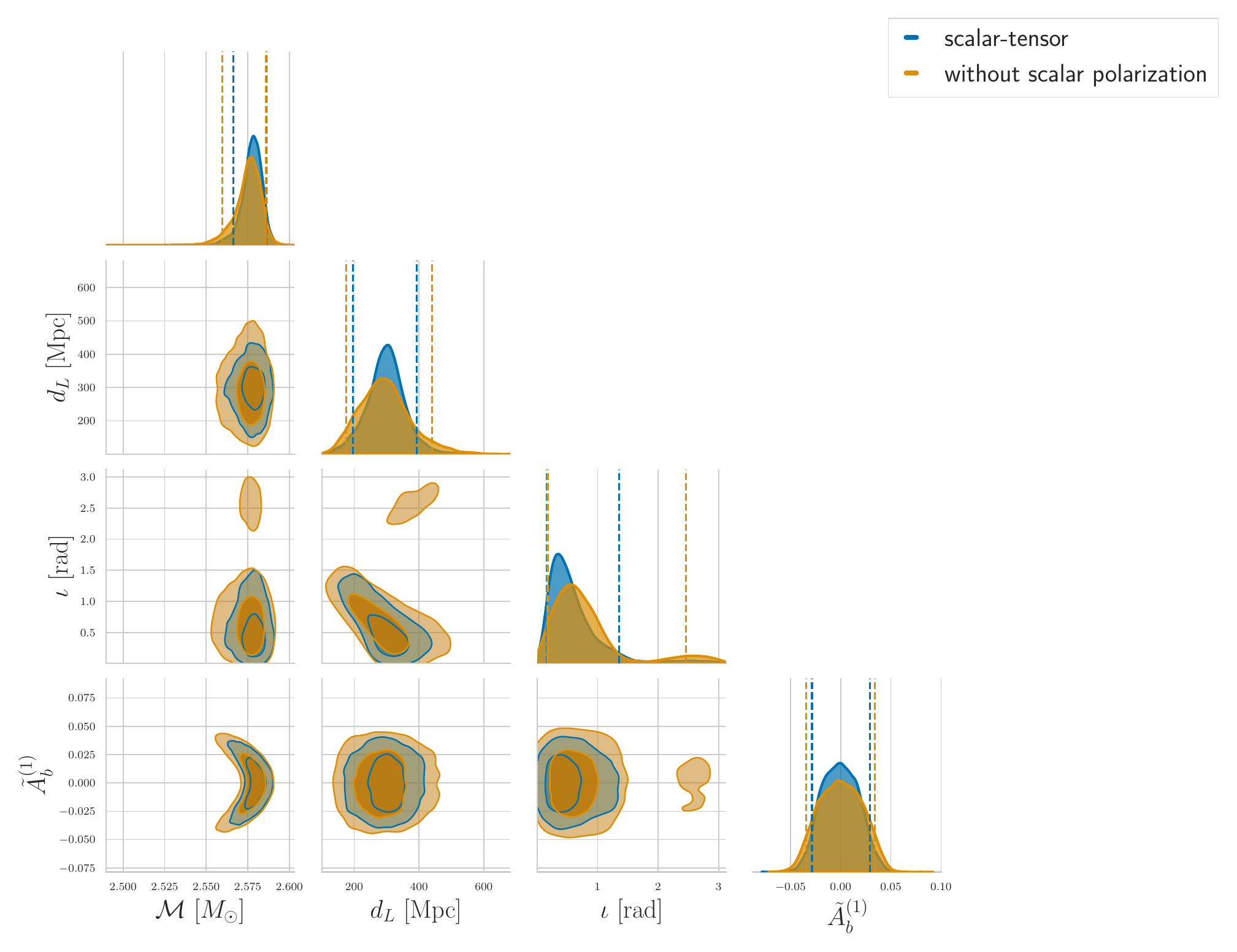}
\end{center}
\caption{This is a similar figure to Fig.~\ref{fig:corner_amplitude_widest} but compared to the analysis using the model of tensor modes with scalar phase corrections.}
\label{fig:corner_amplitude_comparison_without_scalar}
\end{figure}

By implementing scalar modes, the phase parameters appear 
to be estimated differently. The estimated mean value is $|\tilde{A}^{(1)}_{b}|\leq 0.034$ with 90\% credible level, which can be converted into $|\hat{\alpha}_A|\leq 0.049$ through Eq.~\eqref{eq:parms_relations2}. Thus, we expect that the inclusion of the scalar polarization modes would break the parameter degeneracy. Figure~\ref{fig:corner_amplitude_comparison_without_scalar} shows the comparison of the posterior samples for the amplitude parameters between the scalar-tensor model (blue) and the model of tensor modes with scalar corrections (orange). We note that the posterior distribution for the inclination angle $\iota$ changes by including the scalar modes and the gradual peak around $\iota\sim2.5\ {\rm rad}$ has disappeared. We can also find the similar two peaks in the GR analysis as shown in orange in Fig.~\ref{fig:corner_amplitude_widest}. The inclination-angle dependence of the GW radiation differs among the polarization modes~\cite{Takeda:2020tjj}. The overall inclination-angle dependence of the tensor modes is given by $ \sqrt{(1+\cos^2{\iota})^2+4\cos^2{\iota}}$, which has similar values at the positions of the two peaks $\iota\sim 0.5\ {\rm rad}$ and $\iota\sim 2.5\ {\rm rad}$.  However, since the dipole scalar mode is proportional to $\sin{\iota}$, the sign of the amplitude flips between the two peaks. Thus, the inclination dependence of the dipole scalar mode in the waveform model would be helpful to break the partial degeneracy in the inclination angle. It would result in the better constraints on the other parameters.\\

%%%%%%%%%%%%%%%%%%%%%%%%%%%%%%%%%%%%%%%%%%%%%%%%%%%%%%%%%%%%%%%%%%%%%%%%%%
%%%%%%%%%%%%%%%%%%%%%%%%%%%%%%%%%%%%%%%%%%%%%%%%%%%%%%%%%%%%%%%%%%%%%%%%%%%

%%%%%%%%%%%%%%%%%%%%%%%%%%%%%%%%%%%%%%%%%%%%%%%%%%%%%%%%%%%%%%%%%%%%%%%%%%
%%%%%%%%%%%%%%%%%%%%%%%%%%%%%%%%%%%%%%%%%%%%%%%%%%%%%%%%%%%%%%%%%%%%%%%%%%

\paragraph{Tensor modes plus scalar dipole mode}
Second, we perform the Bayesian analysis based on the waveform model including the tensor modes $\tilde{h}_{+,\times}$, dipole scalar mode $\tilde{h}_{b1}$, and the amplitude and phase corrections due to the dipole scalar radiation.
Hence, the waveform model is given by
\be
\tilde{h}_I(f)=\tilde{h}_{T}(f)+\tilde{h}^{(1)}_{b}(f)\,,
\ee
where the GW polarizations are
\ba
\tilde{h}_{T}(f) 
&=& -\left[F_I^{+}(1+\cos^2{\iota})-2iF_I^{\times}\cos{\iota}\right] 
\left[ 1+\delta A^{(2)} \right] 
\sqrt{\frac{5\pi}{96}}\frac{(G_* {{\cal M}})^2}{d_L}
\left( u_{*}^{(2)} \right)^{-7/2}
e^{-i\Psi_\text{GR}^{(2)}}e^{-i\delta\Psi^{(2)}}\,,\\
\tilde{h}^{(1)}_{b}(f)
&=& \sqrt{\frac{5\pi}{48}}A^{(1)}_{b}F_{I}^{b}(2\sin{\iota})
\eta^{1/5}\frac{(G_*{\cal M})^2}{d_L} 
\left( u_{*}^{(1)} \right)^{-9/2} 
e^{-i\Psi_\text{GR}^{(1)}}e^{-i\delta\Psi^{(1)}}\,,
\ea
with the phase corrections
\begin{align}
\delta A ^{(\ell)} =-\frac{5}{48} \left(\tilde{A}_{b}^{(1)} \right)^2 
\eta^{2/5} \left(u_{*}^{(\ell)} \right)^{-2}
\,, \\
\delta \Psi^{(\ell)} =-\frac{5\ell}{3584} 
\left( \tilde{A}_{b}^{(1)} \right)^2
\eta^{2/5} \left( u_{*}^{(\ell)} \right)^{-7} \;.
\end{align}

\begin{figure}[h]
\begin{center}
\includegraphics[height=6.5in,width=6.8in]{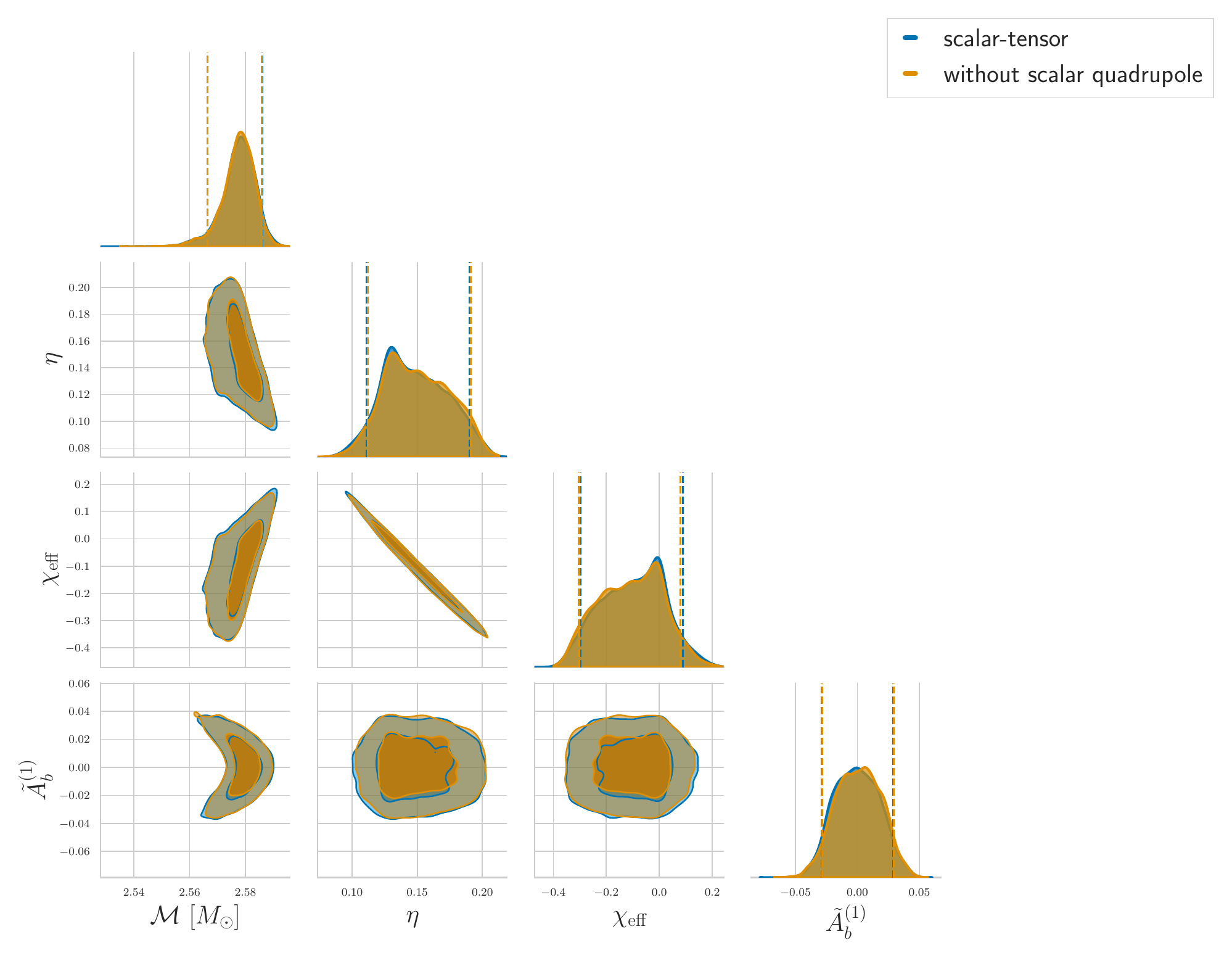}
\end{center}
\caption{This is a similar figure to Fig.~\ref{fig:corner_phase_widest} but compared to the analysis using the model of tensor modes plus scalar dipole mode. }
\label{fig:corner_phase_comparison_without_quadrupole}
\end{figure}

\begin{figure}[ht]
\begin{center}
\includegraphics[height=6.5in,width=6.8in]{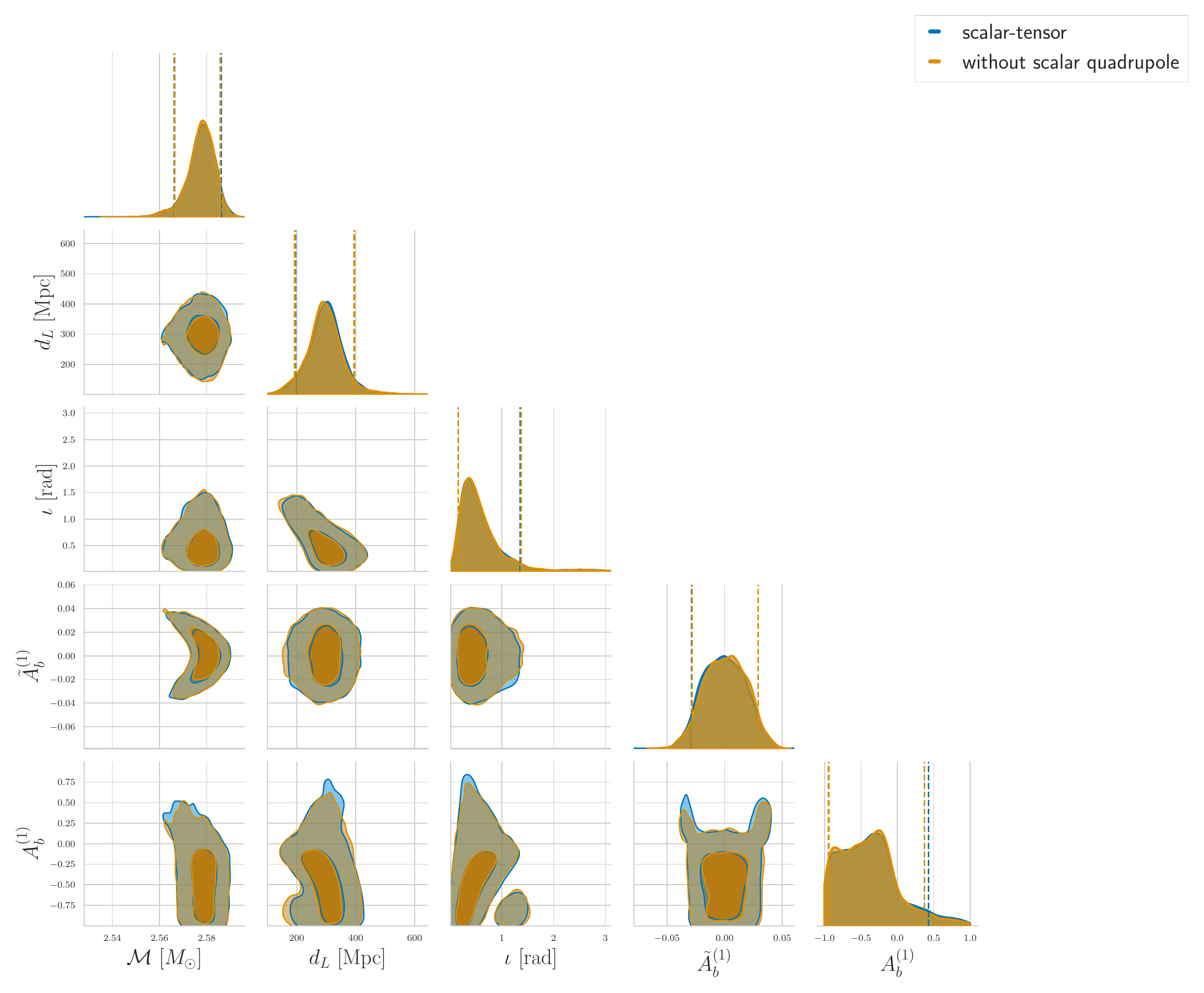}
\end{center}
\caption{This is a similar figure to Fig.~\ref{fig:corner_amplitude_widest} but compared to the analysis using the model of tensor modes plus scalar dipole mode. }
\label{fig:corner_amplitude_comparison_without_quadrupole}
\end{figure}

In this model, we omit the quadrupole scalar radiation because we expect that the dipole radiation becomes dominant at the early inspiral stage. Hence, we can evaluate the contribution of the scalar quadrupole mode to the parameter estimation by comparing the result of this analysis with that using the scalar-tensor model. Fig.~\ref{fig:corner_phase_comparison_without_quadrupole} shows the comparison of the posterior samples for the phase parameters and Fig.~\ref{fig:corner_amplitude_comparison_without_quadrupole} shows the comparison of the posterior samples for the amplitude parameters between the scalar-tensor model (blue) and the model of tensor modes plus scalar dipole mode (orange). The posterior distributions for both are almost identical. This indicates that the results of the current analysis in scalar-tensor are mostly determined by the corrections due to scalar dipole radiation.

%%%%%%%%%%%%%%%%%%%%%%%%%%%%%%%%%%%%%%%%%%%%%%%%%%%%%%%%%%%%%%%%%%%%%%%%%%
%%%%%%%%%%%%%%%%%%%%%%%%%%%%%%%%%%%%%%%%%%%%%%%%%%%%%%%%%%%%%%%%%%%%%%%%%%%

%%%%%%%%%%%%%%%%%%%%%%%%%%%%%%%%%%%%%%%%%%%%%%%%%%%%%%%%%%%%%%%%%%%%%%%%%%
%%%%%%%%%%%%%%%%%%%%%%%%%%%%%%%%%%%%%%%%%%%%%%%%%%%%%%%%%%%%%%%%%%%%%%%%%%

\subsubsection{Scalar-to-tensor amplitude ratio}
We define scalar-to-tensor amplitude ratios as the ratios of scalar-mode amplitude to tensor-mode amplitude, that is, 
from Eqs.~(\ref{eq:hT_parameterized})-(\ref{eq:freq_b2_parameterized}),
\begin{equation}
\label{eq:def_scalar_to_tensor_ratio}
R^{(1)}_{ST}\equiv\left|\frac{A^{(1)}_{b} \eta^{1/5} \sin \iota}{\sqrt{(1+\cos^2 \iota)^2+4\cos^2 \iota}}(G_* \mathcal{M}\pi f)^{-1/3} \right|\,,
\end{equation}
for the scalar dipole mode and
\begin{equation}
R^{(2)}_{ST}\equiv\left|\frac{A^{(2)}_{b} 2\sin^2 \iota}{\sqrt{(1+\cos^2 \iota)^2+4\cos^2 \iota}}\right|\,,
\end{equation}
for the scalar quadrupole mode.
This ratio is an observational indicator to represent how deep the scalar mode is explored  by the GW observation compared to the tensor modes.
As shown in Section \ref{sec:results}, since the contribution of the dipole scalar mode is dominant and the dipole mode amplitude and the quadrupole mode are characterized by the same parameter $A^{(1)}_{b}$, we evaluate the dipole scalar-to-tensor amplitude ratio here. On substituting the estimated mean values and the 90\% credible intervals for the parameters $\mathcal{M}$, $\iota$, $A^{(1)}_{b}$ and the typical GW frequency $f\sim 100\ {\rm Hz}$,
we find the constraints on the ratio as
\begin{align}
\label{eq:constraint_scalar_to_tensor_ratio}
R^{(1)}_{\rm ST} \lesssim 0.57\,,
\end{align}
for GW200115. This scalar-to-tensor amplitude ratio is relatively large compared to the ratios for GW170814 and GW170817 reported in~\cite{Takeda:2021hgo}. As the values of $|\sin \iota|$ and $|\cos \iota|$ are similar to those of the previous events, the reason is because the authors assumed that the coupling $\gamma$ in Eq.~\eqref{eq:parameterized_flux} is unity, and then the scalar-mode amplitude is determined by the phase evolution of the tensor modes. On the other hand, the constraint \eqref{eq:constraint_scalar_to_tensor_ratio} is purely determined by the ability of the GW detector network to probe into the scalar mode. Therefore, this constraint on the scalar-to-tensor amplitude ratio means that the current GW detector network is able to probe into the scalar mode at a level that is at most slightly better than the amplitude of the tensor modes.

%%%%%%%%%%%%%%%%%%%%%%%%%%%%%%%%%%%%%%%%%%%%%
\section{Theoretical interpretation of 
GW200115 constraints}
\label{theoinsec}
%%%%%%%%%%%%%%%%%%%%%%%%%%%%%%%%%%%%%%%%%%%%%

Let us interpret the observational bounds (\ref{aAbound}) and 
(\ref{xi0bound}) as the constraints on model parameters of 
theories encompassed by the action (\ref{action}). 
We will consider two theories: 
(A) BD theories with the
functions (\ref{model1}), and 
(B) theories of spontaneous scalarization of NSs 
with the functions (\ref{model2}). 

For this purpose, we first revisit how $\hat{\alpha}_A$ is 
approximately related to the nonminimal coupling and 
the NS EOS. On a static and spherically symmetric 
background, the large-distance solution far outside the 
NS is given by Eq.~(\ref{phiL}), where $q_s$ is related 
to $\hat{\alpha}_A$ according to Eq.~(\ref{alphaIhat}). 
In the Jordan frame, the radial distance $r$ and the NS ADM 
mass $m_A$ can be expressed as 
$r=\hat{r}/\sqrt{F(\phi)}$ and 
$m_A=\hat{m}_A \sqrt{F(\phi)}$, 
respectively, where $F(\phi)$ is the nonminimal coupling. 
Then, the scalar-field solution 
at large distances is given by 
\be
\phi(r)=\phi_0-\frac{m_A \hat{\alpha}_A}{4\pi \Mpl F(\phi_0)} 
\frac{1}{r}\,.
\label{phiout}
\ee
The scalar-field solution expanded around $r=0$ 
up to the order of $r^2$ is \cite{Higashino:2022izi}
\be
\phi(r)=\phi_c-\frac{\xi_c \rho_c}
{12F(\phi_c)\Mpl} (1-3w_c)r^2\,,
\label{phiin}
\ee
where $\phi_c$ and $\rho_c$ are the field value and the matter density 
at $r=0$, respectively, and $w_c=P_c/\rho_c$ is the EOS  
parameter at $r=0$ with the central pressure $P_c$, and 
\be
\xi_c \equiv \frac{\Mpl F_{,\phi}}{F}\biggl|_{\phi=\phi_c}\,.
\ee
The solution (\ref{phiin}) loses its validity around the NS 
surface ($r=r_s$), but we may extrapolate this solution 
up to $r=r_s$. Extrapolating also the large-distance 
solution (\ref{phiout}) 
down to $r=r_s$ and matching its radial derivative with that of 
Eq.~(\ref{phiin}), we obtain the relation 
\be
\hat{\alpha}_A \simeq -\frac{\xi_c}{2} (1-3w_c)\,,
\label{aIes}
\ee
where we used the approximations $F(\phi_c) \simeq F(\phi_0)$ and 
$m_I \simeq 4\pi \rho_c r_s^3/3$.
The formula (\ref{aIes}) is a crude estimation, 
as it does not incorporate the property of solutions 
in the intermediate regime. Moreover, there may be 
some nonperturbative effects on the solutions 
inside the NS.
However, the estimation (\ref{aIes}) is useful to 
understand what determines the scalar charge physically.
Not only the nonminimal coupling strength $\xi_c$ 
but also the NS 
EOS $w_c$ affects the amplitude of $\hat{\alpha}_I$. 
As we approach $r=r_s$ from $r=0$, the matter EOS
parameter $w=P/\rho$ (with pressure $P$ and density $\rho$) 
decreases toward 0. This means that, for $w_c<1/3$, the 
formula (\ref{aIes}) can underestimate the amplitude of 
$\hat{\alpha}_I$. Moreover, the $\mu(\phi)X^2$ term 
in $G_2$ does not affect the expansion (\ref{phiin}) 
up to the order of $r^2$ by reflecting the fact that 
the leading-order field derivative around $r=0$ is 
the term linear in $X$ (i.e., $\phi'(r) \propto r$). 
However, as $r$ approaches $r_s$, 
we cannot neglect the higher-order term $\mu(\phi)X^2$ 
relative to $X$. This effect modifies the solution to 
the scalar field especially around $r=r_s$.
When $\mu(\phi)$ is negative, there is a kinetic 
screening of $\phi'(r)$, which leads to the suppression 
of $|\hat{\alpha}_A|$ \cite{Higashino:2022izi}.

\subsection{BD theories}

In BD theories, the nonminimal coupling is given by 
$F(\phi)=e^{-2Q \phi/\Mpl}$. In this case,  
the quantities $\xi_0$ and $\xi_c$ reduce to 
\be
\xi_0=\xi_c=-2Q\,,
\ee
which do not depend on $\phi$. 
The bound (\ref{xi0bound}) translates to 
\be
-134 \le Q \le 138\,,
\ee
which is weak due to the poorly constrained value 
of the nonminimal coupling strength $\xi_0$.

On using the approximate relation (\ref{aIes}), we have 
\be
\hat{\alpha}_A \simeq Q (1-3w_c)\,.
\label{halA}
\ee
Applying the bound (\ref{aAbound}) to this approximate relation, 
the coupling constant $Q$ is constrained to be 
\be
|Q| \lesssim \frac{0.04}{1-3w_c}\,.
\label{Qcon0}
\ee
If we take the EOS parameter $w_c=0.2$ as 
a typical value, the bound (\ref{Qcon0}) 
crudely translates to $|Q| \lesssim 0.1$. 

\begin{figure}[h]
\begin{center}
\includegraphics[height=3.2in,width=3.4in]{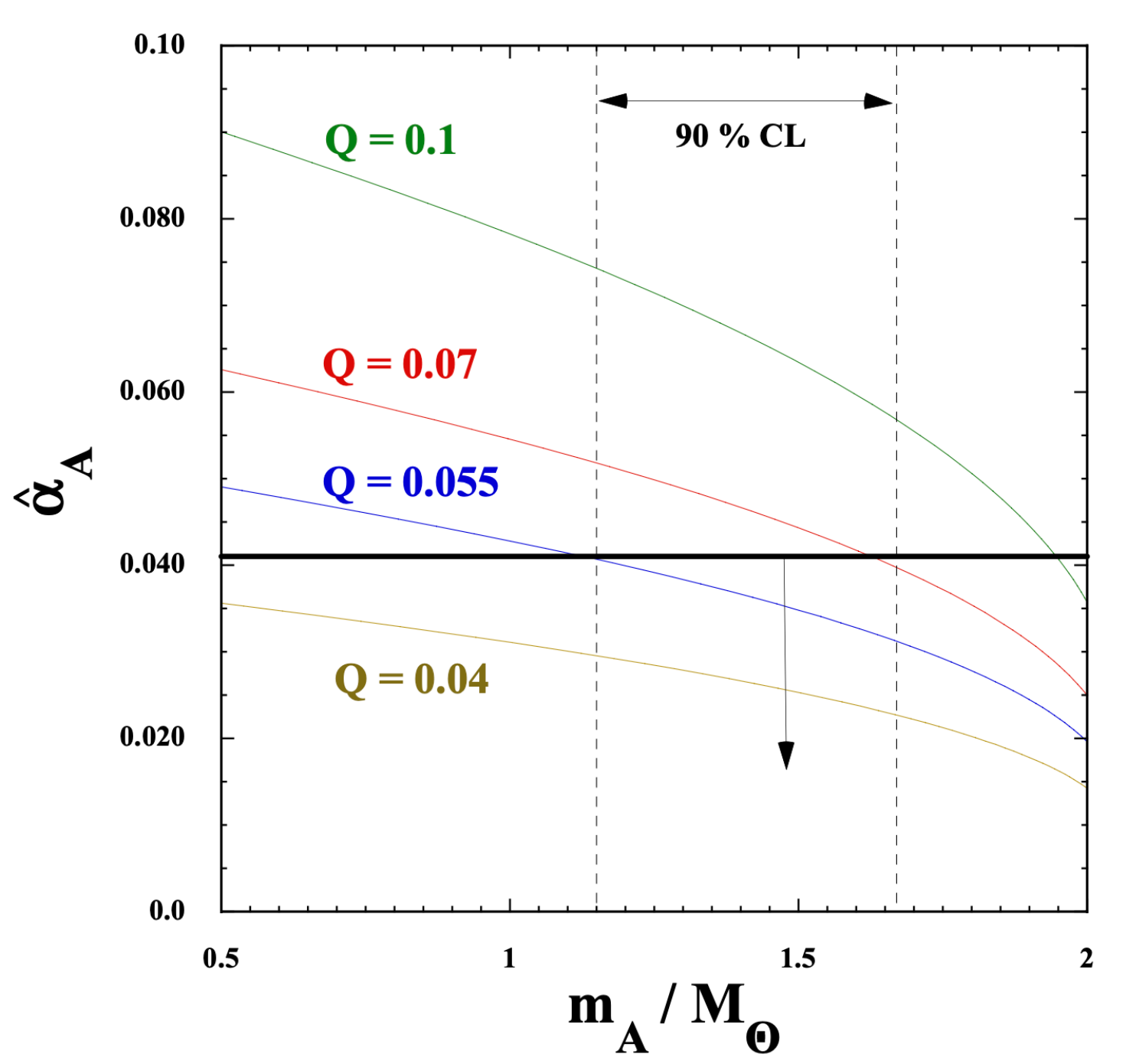}
\end{center}
\caption{The scalar charge $\hat{\alpha}_A$ versus the 
NS mass $m_A$ (normalized by the solar mass $M_{\odot}$) 
for the SLy EOS in BD theories with four different couplings $Q$. 
We also show the upper limit $\hat{\alpha}_A=0.041$ 
constrained from the GW200115 data as a bold black line. 
The NS mass range at 90\,\% credible level shown with the dashed 
vertical lines is $1.15\,M_{\odot} < m_A < 1.67\,M_{\odot}$.
Since the theoretical curves need to be in the ranges 
$\hat{\alpha}_A \le 0.041$ and 
$1.15\,M_{\odot} \le m_A \le 1.67\,M_{\odot}$, 
the coupling $Q$ is bounded as $Q \le 0.055$.
}
\label{fig:BD}
\end{figure}

Since we do not know the precise NS EOS of the GW200115 event, 
there is an uncertainty of the upper limit on $|Q|$. 
By specifying a particular EOS, $\hat{\alpha}_A$ 
can be numerically computed without resorting to 
the approximate formula (\ref{halA}). 
We use an analytic representation of the SLy EOS 
given in Ref.~\cite{Haensel:2004nu}.
We numerically calculate $\hat{\alpha}_A$ by changing the values 
of $Q$ and the central matter density $\rho_c$. 
In Fig.~\ref{fig:BD}, we plot $\hat{\alpha}_A$ versus the 
NS mass $m_A$ (normalized by $M_{\odot}$) for four 
positive values of $Q$. 
Since $m_A$ tends to increase for larger $w_c$ 
in the range $w_c<1/3$, the approximate relation (\ref{halA}) 
implies that $\hat{\alpha}_A$ should decrease as a function of $m_A$. 
Indeed, this behavior of $\hat{\alpha}_A$ versus $m_A$ can be 
confirmed in Fig.~\ref{fig:BD}. 
We note that, for $w_c$ exceeding $1/3$, the approximate 
formula (\ref{halA}) loses its validity. 
Even in such cases, the numerical values of 
$\hat{\alpha}_A$ are typically positive for $Q>0$.

In Fig.~\ref{fig:BD}, we show the GW200115 bound 
$\hat{\alpha}_A \le 0.041$ besides 
the region of $m_A$ constrained from the data, i.e., 
$1.15 M_{\odot} \le m_A \le 1.67 M_{\odot}$.
Requiring that the theoretical curves are within the region 
constrained from the GW 200115 data, we obtain the bound 
\be
|Q| \le 0.055\,,\qquad {\rm or} \qquad 
\omega_{\rm BD} \ge 81\,\qquad 
{\rm for~~SLy~EOS}\,,
\label{QSLy}
\ee
This is tighter than the crude estimation $|Q| \lesssim 0.1$ 
explained above. 
If we choose the NS EOSs other than the SLy, 
the upper limits on $|Q|$ are generally different from 
the bound (\ref{QSLy}). Nevertheless, we expect that $|Q|$ 
does not exceed the order 0.1 as estimated from (\ref{Qcon0}).
In the analysis of tensor modes alone without 
the breathing scalar polarization, which is similar 
to the analysis performed 
in Refs.~\cite{LIGOScientific:2021sio, Niu:2021nic}, we find that 
the constraint on the scalar charge becomes looser, i.e., 
$|\hat{\alpha}_A| \le 0.049$. 
In this case, the bound on the BD parameter is consistent with the 
limit $\omega_{\rm BD} \gtrsim 40$ derived in Ref.~\cite{Niu:2021nic}.
The breaking of the partial parameter degeneracy by adding the scalar mode 
would contribute to the tighter bound.
We recall that, even for $\hat{\alpha}_A<0$, the observational 
upper limit of $|\hat{\alpha}_A|$ in Eq.~(\ref{aAbound}) 
is similar to that for $\hat{\alpha}_A>0$. 
Hence, for $Q<0$, we also obtain the 
bound on $|Q|$ similar to (\ref{QSLy}). 
Although they are weaker than the limit 
$|Q| \le 2.5 \times 10^{-3}$ \cite{Bertotti:2003rm,Khoury:2003rn,Tsujikawa:2008uc,DeFelice:2010aj}
constrained from the solar-system tests by one order of magnitude, 
it is expected that future high-precision GW observations  
can put tighter bounds on $|Q|$.

%%%%%%%%%%%%%%%%%%%%%%%%%%%%%%%%%%%%%%%%%%%%%%%%%%%%%%%%%%%%%%%%%%%%%%%%%%
%%%%%%%%%%%%%%%%%%%%%%%%%%%%%%%%%%%%%%%%%%%%%%%%%%%%%%%%%%%%%%%%%%%%%%%%%%

%
\subsection{Theories of spontaneous scalarization of NSs}

In theories of NS spontaneous scalarization with the nonminimal 
coupling $F(\phi)=e^{-\beta \phi^2/(2\Mpl^2)}$, 
the nonminimal coupling strength far outside the NS 
is given by 
\be
\xi_0=-\frac{\beta \phi_0}{\Mpl}\,.
\ee
Then, the bound (\ref{xi0bound}) translates to 
\be
-268 \Mpl \le \beta \phi_0 \le 276 \Mpl\,.
\label{bphi0} 
\ee
In current theories, the  parameterized post-Newtonian (PPN) 
parameter $\gamma_{\rm PPN}$ 
is given by $\gamma_{\rm PPN}-1=-2\xi_0^2/(2+\xi_0^2)$ \cite{Damour:1992we,Damour:1996ke}.
Then, the solar-system constraint $\gamma_{\rm PPN}-1=(2.1 \pm 2.3) 
\times 10^{-5}$ \cite{Will:2014kxa} translates to 
$|\beta \phi_0| \le 1.4 \times 10^{-3}\Mpl$.
The current limit (\ref{bphi0}) arising from the amplitude of 
scalar GWs is much weaker than the solar-system bound.

\subsubsection{DEF model}

Let us first consider the original DEF model with $\mu(\phi)=0$ 
in the coupling functions (\ref{model2}). 
In this case, we can crudely estimate the scalar charge 
by using Eq.~(\ref{aIes}) as 
\be
\hat{\alpha}_A \simeq \frac{\beta \phi_c}{2\Mpl} 
(1-3w_c)\,,
\label{haAes}
\ee
which depends on the field value $\phi_c$ and 
the NS EOS $w_c$ around $r=0$. 
Spontaneous scalarization of NSs occurs for 
$\beta \le -4.35$ due to the tachyonic instability of 
the $\phi=0$ GR branch\footnote{For the matter EOS 
$w_c>1/3$, it is also possible to realize spontaneous scalarization even with positive couplings 
$\beta$ \cite{Palenzuela:2015ima,Mendes:2016fby,Doneva:2022ewd}. 
In the original DEF model, however, scalarized NS 
solutions can be unstable as compared to the GR branch 
for $\beta \gg 1$. We will not consider such a 
positive value of $\beta$.}. 
This critical value of $\beta$ is insensitive to the choices 
of NS EOSs \cite{Harada:1998ge,Novak:1998rk,Silva:2014fca,Barausse:2012da}.
When spontaneous scalarization occurs, the field value $\phi_c$ 
can be as large as ${\cal O}(0.1 \Mpl)$. 
Then, the estimation (\ref{haAes}) shows that $|\hat{\alpha}_A|$ 
should reach the order $0.1$. 
Since the amplitude of $\hat{\alpha}_A$ depends on 
the coupling constant $\beta$, it is possible to put 
constraints on $\beta$ by using the observational bound (\ref{aAbound}). 
The scalar charge $\hat{\alpha}_A$ also carries the information 
of NS EOSs, but, for given $\beta$, the maximum values of 
$|\hat{\alpha}_A|$ weakly depend on the choices of NS EOSs \cite{Harada:1998ge,Novak:1998rk,Silva:2014fca,Barausse:2012da,Niu:2021nic} (see also Ref.~\cite{Shibata:2013pra} 
for dynamical scalarization of NSs during the 
final stage of the inspiral).

For the SLy EOS, we numerically compute $\hat{\alpha}_A$ 
by choosing several different values of $\beta$. 
For given $\beta$ and $\rho_c$, we will iteratively find a 
boundary value of $\phi_c$ leading to a scalarized solution 
with the asymptotic field value
$\phi_0$ close to 0 (to be consistent with the solar-system 
bound mentioned above). 
Different choices of $\rho_c$ lead to different NS masses and 
scalar charges. Spontaneous scalarization of NSs occurs for 
intermediate central densities $\rho_c$, in which regime 
$\hat{\alpha}_A$ is nonvanishing. 

%%%%%%%%%%%%%%%%%%%%%%%%%%%%%%%%%%%%%%%%%%%%%%%%%%%%%%%%%%%%%%%%%%%%%%%%%%
\begin{figure}[h]
\begin{center}
\includegraphics[height=3.0in,width=3.2in]{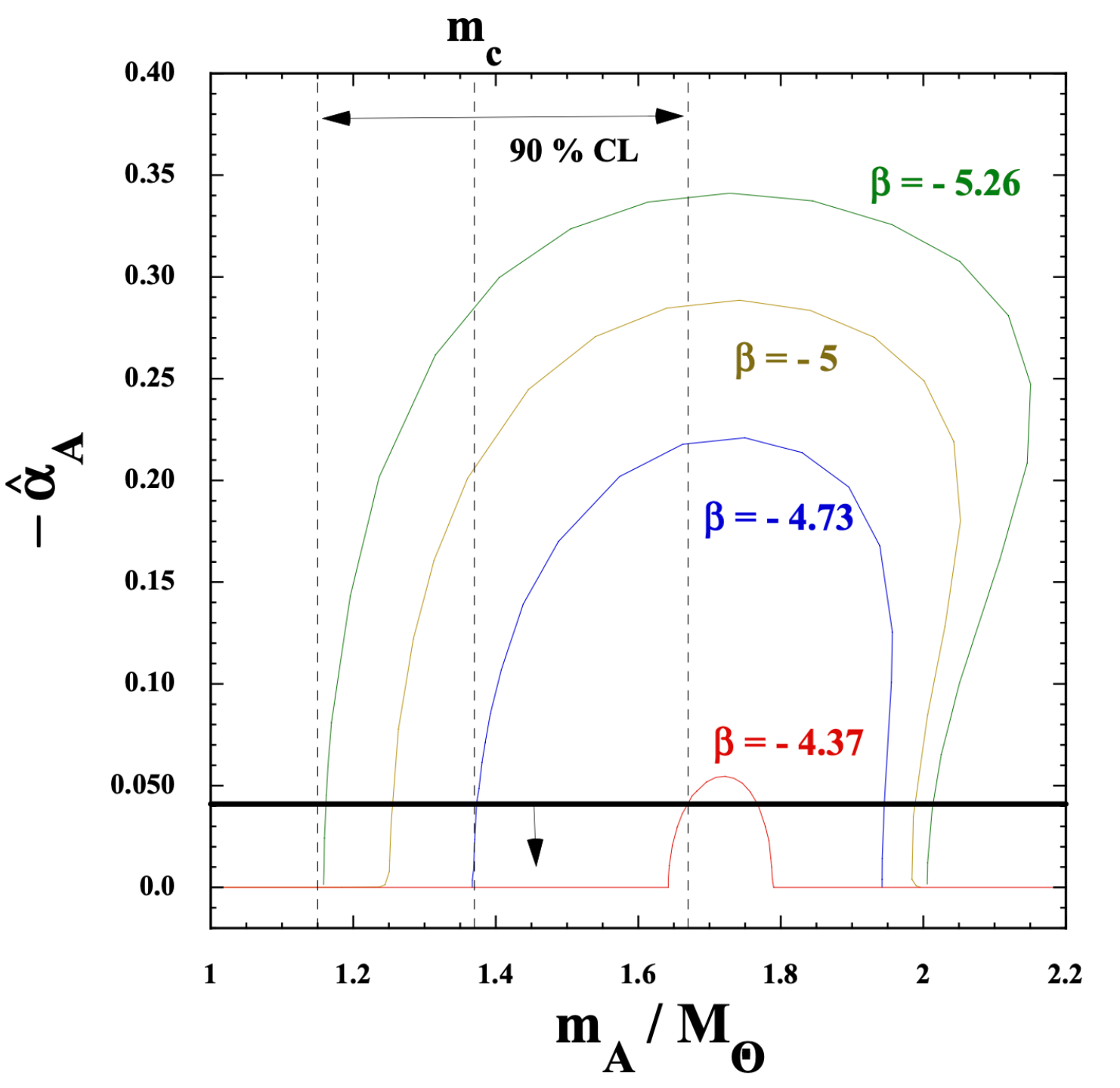}
\end{center}
\caption{We plot $-\hat{\alpha}_A$ versus the NS mass $m_A$ 
(normalized by the solar mass $M_{\odot}$) for the SLy EOS 
in original DEF theories of spontaneous scalarization.
We choose four different coupling constants: 
$\beta=-5.26, -5, -4.73, -4.37$. 
The upper limit $-\hat{\alpha}_A=0.041$ 
constrained from the GW200115 data is 
plotted as a bold black line. 
The NS mass range at 90\,\% credible level shown with the dashed 
vertical lines is $1.15\,M_{\odot} \le m_A \le 1.67\,M_{\odot}$.
The observationally allowed region corresponds to
a rectangle characterized by $0 \le -\hat{\alpha}_A \le 0.041$
and $1.15\,M_{\odot} \le m_A \le 1.67\,M_{\odot}$.
}
\label{fig:spo1}
\end{figure}
%%%%%%%%%%%%%%%%%%%%%%%%%%%%%%%%%%%%%%%%%%%%%%%%%%%%%%%%%%%%%%%%%%%%%%%%%%

In Fig.~\ref{fig:spo1}, we show $-\hat{\alpha}_A$ versus 
$m_A/M_{\odot}$ for four different values of $\beta$ 
with the choice of the SLy EOS.
Since $\beta<0$ for the occurence of scalarization, 
$\hat{\alpha}_A$ is negative if $\phi_c>0$. 
For $\beta=-5.26$, spontaneous scalarization 
occurs for the mass range $1.16 M_{\odot} \le m_A \le 2.01 M_{\odot}$. 
In the rest of the mass region, the scalar charge 
is vanishing ($\hat{\alpha}_A=0$). 
For $\beta<-5.26$, the theoretical curve is outside 
the observationally allowed region ($0 \le -\hat{\alpha}_A \le 0.041$) 
for any NS mass constrained from the GW200115 event
($1.15M_{\odot} \le m_A \le 1.67M_{\odot}$). 
Then, we obtain the bound 
\be
\beta \ge -5.26\,,
\label{betabo}
\ee
for the SLy EOS. 
As $\beta$ increases, the mass region in which scalarization
takes places gets narrower, with smaller maximum values of 
$|\hat{\alpha}_A|$. 
For $-5.26 \le \beta \le -4.37$, there are the mass ranges 
excluded by the observational limit $|\hat{\alpha}_A| \le 0.041$, 
together with the existence of allowed 
mass parameters within the mass range of the GW200115 event. 
As $\beta$ increases toward $-4.37$, the observationally excluded 
mass region tends to be narrower. 
In particular, the model with $\beta \ge -4.37$ is consistent 
with the bound on $\hat{\alpha}_A$ for 
all the constrained mass ranges. 
This is much stronger than the constraint (\ref{betabo}), but we 
have to caution that there are still allowed mass regions 
even for $-5.26 \le \beta \le -4.37$. 
In this sense, we will take the conservative 
limit (\ref{betabo}) as the bound on $\beta$ 
constrained from the GW200115 data. 

Clearly, the observational uncertainty of the NS 
mass gives the limitation for putting tight bounds on $\beta$. Let us take the central value of $m_A$ constrained from 
the data, i.e., $m_c=1.37M_{\odot}$. 
If we demand the condition $|\hat{\alpha}_A| \le 0.041$ 
in this case, the coupling needs to be in the range 
\be
\beta \ge -4.73\,,\qquad {\rm for} 
\quad m_A=1.37M_{\odot}\,.
\label{betabo2}
\ee
This shows that, if the NS mass is constrained to be in 
a narrower range, it is possible to place tighter bounds 
on $\beta$ than (\ref{betabo}). Since this can happen 
in future observations, the accumulation of many NS-BH merger 
events will clarify whether the original spontaneous scalarization 
scenario is excluded or not.
Combined with the condition for the occurenece of spontaneous scalarization, 
the coupling $\beta$ is now constrained to be $-5.26 \le \beta \le -4.35$. 
Hence the allowed range of $\beta$ is already narrow even with 
a single GW event.
We note that, in the analysis of Ref.~\cite{Niu:2019ywx}, the authors 
extended the range of $\beta$ to $\beta>-4.35$ in which 
spontaneous scalarization does not occur. 
In this regime, the solution stays in the GR branch 
$\phi(r)=0$ and hence $\hat{\alpha}_A=0$.
In this sense, the constraint 
on $\beta$ in the scalarization scenario is meaningful 
only for $\beta \le -4.35$. 

\subsubsection{DEF model with kinetic screening}

\begin{figure}[h]
\begin{center}
\includegraphics[height=3.0in,width=3.2in]{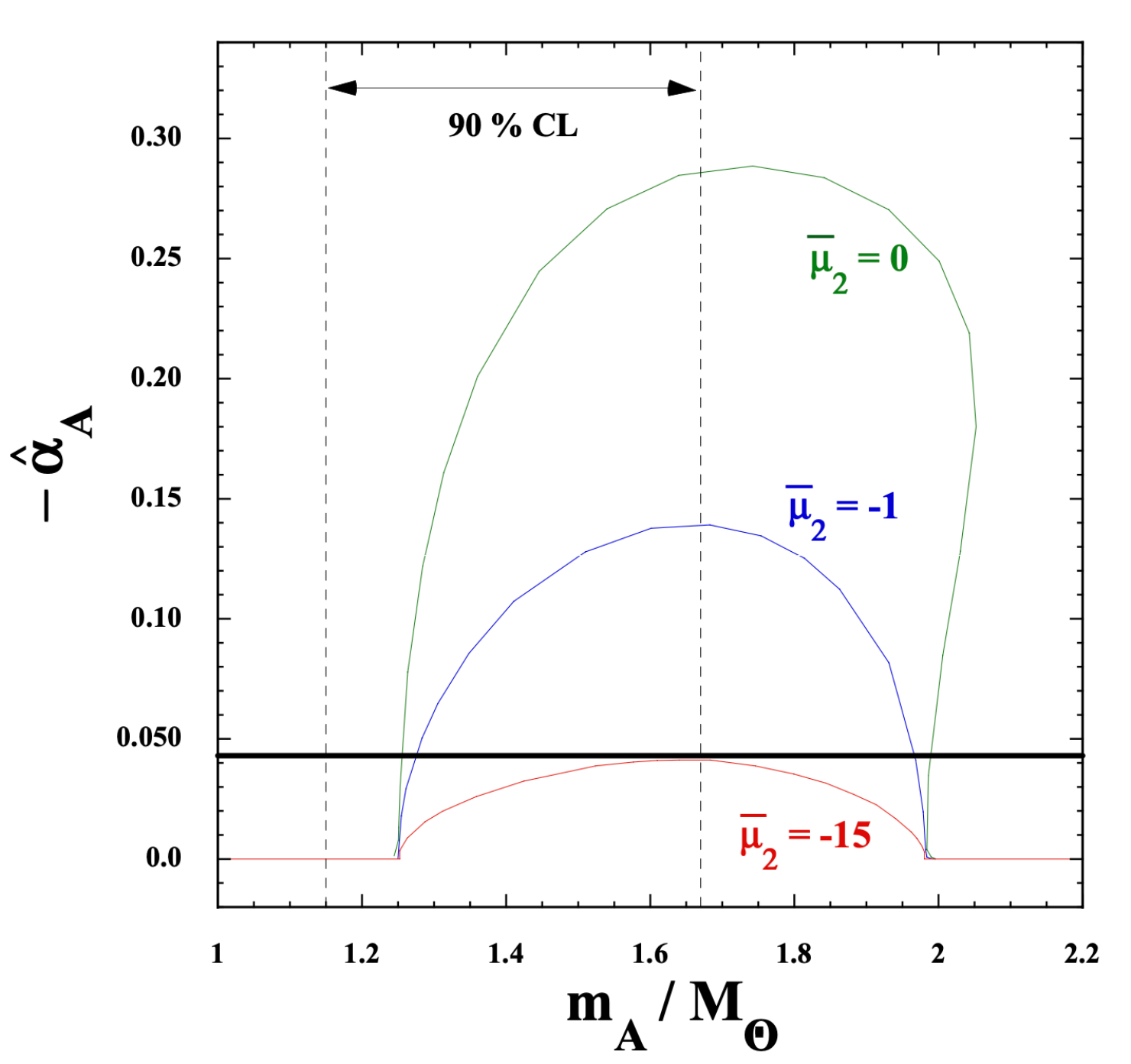}
\end{center}
\caption{$-\hat{\alpha}_A$ versus $m_A/M_{\odot}$
for the SLy EOS in theories of spontaneous scalarization 
with $\beta=-5$ in the presence of the higher-order 
derivative term $\mu_2 X^2$. Each line corresponds to 
$\bar{\mu}_2=0, -1, -15$ from top to bottom, 
where $\bar{\mu}_2=\mu_2 \Mpl^2/r_0^2$ with $r_0=89.664$~km. 
We also show the observational limits of 
$-\hat{\alpha}_A$ and $m_A$. 
For $\bar{\mu}_2 \leq -15$, the theoretical lines are 
within the region constrained from the data.}
\label{fig:spo2}
\end{figure}

Let us next proceed to the case in which there is a term 
$\mu_2 X^2$ in the function $G_2$ of Eq.~(\ref{model2}), 
i.e., $\mu(\phi)=\mu_2={\rm constant}$.  
For $\mu_2<0$, it was shown in Ref.~\cite{Higashino:2022izi} 
that the higher-order derivative term can lead to 
a kinetic screening of the scalar field inside the NS. 
As we see in Fig.~\ref{fig:spo2}, the scalar charge 
$|\hat{\alpha}_A|$ tends to be smaller for decreasing 
values of $\mu_2$. 
The NS mass range in which spontaneous scalarization occurs is 
insensitive to the coupling constant $\mu_2$. 
In other words, the maximum values of $|\hat{\alpha}_A|$ 
get smaller for decreasing $\mu_2$, but the mass range 
with $|\hat{\alpha}_A|>0.01$ is hardly modified. 
This properly is different from the spontaneous scalarization 
scenario with $\mu_2=0$, in which case the mass 
region with $|\hat{\alpha}_A|>0.01$ shrinks for increasing 
$\beta$ (see Fig.~\ref{fig:spo1}).

%%%%%%%%%%%%%%%%%%%%%%%%%%%%%%%%%%%%%%%%%%%%%%%%%%%%%%%%%%%%%%%%%%%%%%%%%%
%%%%%%%%%%%%%%%%%%%%%%%%%%%%%%%%%%%%%%%%%%%%%%%%%%%%%%%%%%%%%%%%%%%%%%%%%%

As an example, let us consider the case where $\beta=-5$. 
Then, the condition (\ref{betabo}) is satisfied in the original 
scalarization scenario, but there are the NS mass ranges in which 
the inequality $|\hat{\alpha}_A| \le 0.041$ is violated. 
Indeed, for the medium mass $m_A=1.37M_{\odot}$, this bound 
on $\hat{\alpha}_A$ is not satisfied, see Eq.~(\ref{betabo2}).
In Fig.~\ref{fig:spo2}, we plot $-\hat{\alpha}_A$ versus $m_A$ 
for $\bar{\mu}_2=0, -1, -15$ by fixing $\beta=-5$, where 
$\bar{\mu}_2 \equiv \mu_2 \Mpl^2/r_0^2$ and $r_0=89.664$~km.  
The theoretical curves are within the observationally 
constrained region of $\hat{\alpha}_A$ and $m_A$ if
\be
\bar{\mu}_2 \le -15\,,
\qquad {\rm for} \quad \beta=-5\,.
\ee
Thus, even for $\beta=-5$, the kinetic screening induced by the higher-order 
derivative term $\mu_2 X^2$ leads to the compatibility with the GW200115 data 
for all the constrained mass ranges.
Thus, for $\mu_2<0$, the allowed parameter region of $\beta$ is 
not restricted to be small unlike the DEF model with $\mu_2=0$.

%%%%%%%%%%%%%%%%%%%%%%%%%%%%%%%%%%%%%%%%%%%%%
\section{Conclusions}
\label{conclusion}
%%%%%%%%%%%%%%%%%%%%%%%%%%%%%%%%%%%%%%%%%%%%%

Table.~\ref{tab:constraints_summary} shows the summary of constraints on the parameters achieved by our analysis of the NS-BH merger event GW200115 in the scalar-tensor framework.

\begin{table}[ht]
\centering
\caption{Summary of constraints on parameters. The credible interval is $90\%$.}
\label{tab:constraints_summary}
\begin{tabular}{@{}lcc@{}}
\toprule
Parameter & Constraint & Notes \\
\midrule
\multicolumn{3}{c}{Analytical Parameters} \\
\midrule
Phase correction parameter: Eq.~\eqref{eq:phase_correction}  & $\tilde{A}^{(1)}_{b}=-0.001^{+0.030}_{-0.028}$ &   \\
Scalar-mode amplitude parameter: Eq.~\eqref{eq:freq_b1_parameterized} & $A^{(1)}_{b}=-0.41^{+0.84}_{-0.54}$ &   \\
Scalar-to-tensor amplitude ratio: Eq.~\eqref{eq:def_scalar_to_tensor_ratio} & $R^{(1)}_{\rm ST} \lesssim 0.57$ & \\
\midrule
\multicolumn{3}{c}{Phenomenological Parameters} \\
\midrule
Scalar charge in Einstein frame: Eq.~\eqref{alphaIhat} & $\hat{\alpha}_A=-0.001^{+0.042}_{-0.040}$ & See Eq.~\eqref{eq:parms_relations2} for the relation $\tilde{A}^{(1)}_{b} = \frac{1}{\sqrt{2}} \hat{\alpha}_A$\\
Nonminimal coupling parameter: Eq.~\eqref{xi_0} & $\xi_{0} = 3.10^{+265}_{-279}$ & See Eq.~\eqref{eq:parms_relations} for the relation $A^{(1)}_{b} =\frac{1}{2}\xi_0 \hat{\alpha}_A$.\\
\midrule
\multicolumn{3}{c}{Theoretical Parameters} \\
\midrule
$Q$ (in BD theories) & $-134 \le Q \le 138$ & Derived from constraint on $\xi_0$ alone.\\
$Q$ (in BD theories) & $|Q| \le 0.055$ & Derived from constraint on $\hat{\alpha}_A$ for SLy EoS.\\
$\beta \phi_0$ (in spontaneous scalarization) 
& $-268 \Mpl \le \beta \phi_0 \le 276 \Mpl$ & cf. $|\beta \phi_0| \le 1.4\times 10^{-3} \Mpl$ (solar system experiment) \\
$\beta$ (in DEF model of spontaneous scalarization) & $\beta \ge -5.26$ & For SLy EoS.\\
\bottomrule
\end{tabular}
\end{table}

From the NS-BH merger event GW200115, we have placed observational 
constraints on the NS scalar charge and the nonminimal coupling 
strength. For this purpose, we chose a subclass of Horndeski theories 
given by the action (\ref{action}), in which case the speed of 
gravity is luminal. In such theories the BHs are 
not endowed with scalar hairs, but it is possible for 
the NSs to have hairy solutions through a scalar-matter 
interaction induced by the nonminimal coupling $G_4(\phi)R$. 
The representative examples are BD theories 
and theories of spontaneous scalarization given by the Horndeski 
functions (\ref{model1}) and (\ref{model2}), respectively.

In theories given by the action (\ref{action}), the inspiral 
gravitational waveforms emitted from compact binaries were 
computed in Ref.~\cite{Higashino:2022izi}. 
In Sec.~\ref{sec:theory}, we reviewed the derivation of time-domain 
GW solutions by assuming that the scalar field is massless. 
In this case, there are two tensor polarizations $h_{+}$ and 
$h_{\times}$ besides a breathing mode $h_b$ arising from the 
scalar-field perturbation. Focusing on a NS-BH binary system 
in which the BH has a vanishing scalar charge, 
we derived the inspiral waveforms of 
frequency-domain GWs propagating on the cosmological background. 
They are given by Eqs.~(\ref{hpFi})-(\ref{hbFi}) with the 
amplitudes (\ref{gwam})-(\ref{hfs2}) and phases 
(\ref{Psipl2})-(\ref{PsibF}). In particular, the breathing mode 
in the frequency domain is newly obtained in this paper. 

In Sec.~\ref{sec:model}, we provided a parameterized 
scalar-tensor inspiral waveform model 
starting with a modified energy flux. Our parameterized waveforms include not only 
the corrections for the tensor modes stemming from scalar radiation 
but also the scalar polarization modes itself. 
Originally, the model includes four additional parameters. 
The two parameters $A_{b}^{(1)}$ and $A_{b}^{(2)}$ represent 
the amplitudes of dipole and quadrupole scalar polarizations, and the two parameters 
$\tilde{A}_{b}^{(1)}$ and $\tilde{A}_{b}^{(2)}$ characterize the waveform 
corrections due to the dipole and quadrupole scalar radiation. 
Since the parameterized waveform model encompasses the theoretical waveforms derived 
in the underlying scalar-tensor theory, we also provided the correspondence between 
the parametrized waveforms and the theoretical waveforms.
According to the correspondence, the two parameters 
$A_{b}^{(1)}$ and $\tilde{A}_{b}^{(1)}$ can be expressed 
in terms of the NS scalar charge 
$\hat{\alpha}_A$ and the nonminimal coupling strength $\xi_0$, as 
$A_{b}^{(1)}=\xi_0 \hat{\alpha}_A/2$ and 
$\tilde{A}_{b}^{(1)}=\hat{\alpha}_A/\sqrt{2}$.
The remaining two parameters $A_{b}^{(2)}$ and $\tilde{A}_{b}^{(2)}$ are 
given by multiplying $A_{b}^{(1)}$ and $\tilde{A}_{b}^{(1)}$ 
by a factor depending on the mass ratio.
Thus, we have two additional free parameters $A_{b}^{(1)}$ and $\tilde{A}_{b}^{(1)}$, 
which can be constrained from the GW observations, in the subclass of Horndeski theories.

In Sec.~\ref{sec:data_analysis}, we placed observational bounds
on the two model parameters $\hat{\alpha}_A$ and $\xi_0$ through 
the constraints on $A_{b}^{(1)}$ and $\tilde{A}_{b}^{(1)}$ 
achieved by the analysis of the GW200115 signal. 
The crucial difference from past works is that we have implemented 
the breathing scalar mode besides the two tensor polarizations.
The inclination angle dependence of the scalar mode may break the partial degeneracy in the estimation of the inclination angle and improve the precision of the determination of the other parameters.
The quantity $\hat{\alpha}_A$ appears in the phases of tensor 
waves at $-1$PN order. This property allows us to put the tight 
bound $\hat{\alpha}_A=-0.001^{+0.042}_{-0.040}$ from the GW200115 data, 
so that the amplitude of the scalar charge is bounded as 
$|\hat{\alpha}_A| \lesssim 0.04$. 
The amplitude of scalar GWs relative to tensor waves can put 
an upper limit on the product $\xi_0 \hat{\alpha}_A$. 
Reflecting the weak observational sensitivity of scalar waves 
in current measurements, 
we obtained the loose bound 
$-1.88<\xi_0 \hat{\alpha}_A<0.86$. 
Combining this with the limit on $\hat{\alpha}_A$, 
the nonminimal coupling strength is also weakly constrained to 
be $\xi_{0}=3.10^{+265}_{-279}$. 
We note that the NS mass is constrained to be in the range 
$1.15 M_{\odot} < m_A < 1.67 M_{\odot}$ from the GW200115 data.

In Sec.~\ref{theoinsec}, we translated the theoretical bounds 
on $\hat{\alpha}_A$ and $\xi_0$ into the constraints on 
model parameters in theories with the Horndeski functions 
(\ref{model1}) and (\ref{model2}). 
In BD theories, the parameter $\xi_0$ is simply equivalent 
to the coupling constant $-2Q$, so that the observational 
constraint on $\xi_0$ alone gives a weak 
limit $-134 \le Q \le 138$.
On the other hand, the other parameter 
$\hat{\alpha}_A$ depends on both $Q$ and the NS EOS.
Extrapolating the solutions around $r=0$ and large distances 
at the NS surface, we obtained the crude formula 
$\hat{\alpha}_A \simeq Q (1-3w_c)$. 
On using the upper limit $|\hat{\alpha}_A| \lesssim 0.04$ and 
taking a typical value $w_c=0.2$, $|\hat{\alpha}_A|$ is smaller 
than the order 0.1. We numerically computed the values 
of $|\hat{\alpha}_A|$ for the SLy EOS without using the 
above approximation and derived the observational 
bound $|Q| \le 0.055$, or equivalently, 
$\omega_{\rm BD} \geq 81$. 

In the DEF model of spontaneous scalarization, 
$|\hat{\alpha}_A|$ can be as large as the order 0.1 
depending on the coupling constant $\beta$. 
Provided that $\beta \ge -5.26$, there are the observationally 
constrained NS mass ranges in which the condition 
$|\hat{\alpha}_A| \le 0.041$ is satisfied. 
Taking the central value of $m_A$ constrained from the data 
($m_A=1.37 M_{\odot}$), the bound on $\hat{\alpha}_A$ 
translates to $\beta \ge -4.73$. 
Due to the uncertainty of the NS mass, we take the most 
conservative bound $\beta \ge -5.26$ as a firm constraint 
extracted from the GW200115 data. 
Since the spontaneous scalarization occurs for $\beta \le -4.35$, 
our newly derived bound already restricts the viable parameter 
space of $\beta$ to a narrow region.
If we take into account the higher-order kinetic term 
$\mu_2 X^2$ with $\mu_2<0$, the kinetic screening mechanism works to reduce 
$|\hat{\alpha}_A|$ without changing the scalarized NS mass region much. 
Taking the nonminimal coupling constant $\beta=-5$, 
the scalarization model with $\mu_2 \Mpl^2/r_0^2 \le -15$ is
compatible with the GW200115 bound for all the constrained 
NS mass ranges.

We have thus shown that the NS-BH merger event GW200115 allows 
us to probe the property of hairy NS solutions. 
In particular, the observational constraint on the scalar charge 
$|\hat{\alpha}_A| \lesssim 0.04$ gives new bounds on the 
nonminimal coupling constants of BD theories and spontaneous
scalarization scenarios with/without the kinetic screening.
Furthermore, our analysis of the parameterized waveforms with
scalar modes indicates that the presence of polarization modes 
beyond GR must be taken into account when attempting to interpret the results 
of parameterized tests with a specific theory of gravity.
Otherwise, observational constraints on alternative theories 
of gravity may be biased.
If the future GW observations were to reach the upper limit of  
$|\hat{\alpha}_A|$ below the order 0.01 with tighter constraints 
on $m_A$, it will be potentially possible to rule out the DEF model. 
Moreover, the increased sensitivity for measuring  
the amplitude of scalar GWs relative to tensor GWs will 
allow us to put tighter constraints on the nonminimal 
coupling strength $\xi_0$~\cite{Takeda:2023mhl}.

%%%%%%%%%%%%%%%%%%%%%%%%%%%%
\section*{Acknowledgements}
%%%%%%%%%%%%%%%%%%%%%%%%%%%%

We would like to thank Hayato Imafuku, Koutarou Kyutoku, Soichiro Morisaki, Rui Niu, Takahiro Tanaka, and Daiki Watarai for useful comments and Kipp Cannon for providing computer usage.
HT is supported by the Grant-in-Aid for Scientific Research 
Fund of the JSPS No.~21J01383 and No.~22K14037.
ST is supported by the Grant-in-Aid for Scientific Research 
Fund of the JSPS No.~22K03642 and Waseda University 
Special Research Project No.~2023C-473. 
AN is supported by JSPS KAKENHI Grant Nos. JP23K03408, JP23H00110, and JP23H04893.
This research has made use of data or software obtained from the Gravitational Wave Open Science Center (gwosc.org), a service of the LIGO Scientific Collaboration, the Virgo Collaboration, and KAGRA. This material is based upon work supported by NSF's LIGO Laboratory which is a major facility fully funded by the National Science Foundation, as well as the Science and Technology Facilities Council (STFC) of the United Kingdom, the Max-Planck-Society (MPS), and the State of Niedersachsen/Germany for support of the construction of Advanced LIGO and construction and operation of the GEO600 detector. Additional support for Advanced LIGO was provided by the Australian Research Council. Virgo is funded, through the European Gravitational Observatory (EGO), by the French Centre National de Recherche Scientifique (CNRS), the Italian Istituto Nazionale di Fisica Nucleare (INFN) and the Dutch Nikhef, with contributions by institutions from Belgium, Germany, Greece, Hungary, Ireland, Japan, Monaco, Poland, Portugal, Spain. KAGRA is supported by Ministry of Education, Culture, Sports, Science and Technology (MEXT), Japan Society for the Promotion of Science (JSPS) in Japan; National Research Foundation (NRF) and Ministry of Science and ICT (MSIT) in Korea; Academia Sinica (AS) and National Science and Technology Council (NSTC) in Taiwan.

\bibliographystyle{mybibstyle}
\bibliography{bib}

\end{document}